\documentclass[journal, twoside]{IEEEtran}

\usepackage{amsmath, amsfonts, amssymb, mathtools, bm}
\usepackage{algorithm, algorithmic}
\usepackage{array, multirow, tabularx, makecell, booktabs}
\usepackage[caption=false,font=normalsize,labelfont=sf,textfont=sf]{subfig}
\usepackage{stfloats, float, wrapfig, threeparttable}
\usepackage{graphicx, graphics, adjustbox}
\usepackage{soul}
\usepackage[dvipsnames, table, xcdraw]{xcolor} 
\usepackage{enumitem, pifont, bbding}
\usepackage{lettrine, textcomp, verbatim, url, cite, hyperref, balance}
\usepackage[utf8]{inputenc}
\usepackage{mathptmx}
\usepackage{orcidlink} 
\usepackage{siunitx}

\usepackage{booktabs}
\usepackage{tabularx}
\usepackage{siunitx} 
\usepackage{colortbl}
\usepackage[table]{xcolor} 
\usepackage{stackengine}

\newcommand{\cmark}{\ding{51}}%
\newcommand{\xmark}{\ding{55}}%

\definecolor{bsRed}{rgb}{0.95, 0.0, 0.0}

\newcommand{\mrev}[1]{{\textcolor{blue}{#1}}}

\begin{document}

\title{CodecFake+: Codec-Based Resynthesized Data as a Proxy for Detecting CodecFake Speech}


\author{
    {Xuanjun Chen$^{*}$}, 
    {Jiawei Du$^{*}$}, 
    {Haibin Wu$^{\dagger}$}, 
    {Lin Zhang}, 
    {I-Ming Lin}, 
    {I-Hsiang Chiu}, 
    {Wenze Ren} 
    \\ {Yuan Tseng}, 
    {Yu Tsao},  
    {Jyh-Shing Roger Jang}, 
    and {Hung-yi Lee} 
\thanks{
This work was supported in part by the National Science and Technology Council, Taiwan, under the Grant 114-2628-E-002-022 and Grant 112-2634-F-002-005. 
This work was supported by the Ministry of Education (MOE) of Taiwan under the project Taiwan Centers of Excellence in Artificial Intelligence, through the NTU Artificial Intelligence Center of Research Excellence (NTU AI-CoRE). \\ 
$^{*}$ denotes Equal contribution. $^{\dagger}$ denotes Corresponding author}
\thanks{Xuanjun Chen, Haibin Wu, Wenze Ren, and Yuan Tseng are with the Graduate Institute of Communication Engineering, National
Taiwan University, Taipei 10617 
(email: \{d12942018, f07921092, r11942166, r11942082\}@ntu.edu.tw)
}
\thanks{Jiawei Du, I-Ming Lin, and Jyh-Shing Roger Jang are with the Department of Computer Science and Information Engineering, National Taiwan University, Taipei 10617 
(email: \{r11922185, r12922130\}@ntu.edu.tw; jang@csie.ntu.edu.tw)
}
\thanks{Lin Zhang is with the Center for Language and Speech Processing at Johns Hopkins University, MD 21218, United States. (e-mail: zlin@ieee.org)}
\thanks{I-Hsiang Chiu is with the Department of Electrical Engineering, National Taiwan University, Taipei 10617 (e-mail: b09901058@ntu.edu.tw)}
\thanks{Yu Tsao is with the Research Center for Information Technology Innovation, Academia Sinica, Taipei 11529 (e-mail: yu.tsao@citi.sinica.edu.tw)}
\thanks{Hung-yi Lee is with the NTU Artificial Intelligence Center of Research Excellence (NTU AI-CoRE), Taipei 10617 (e-mail: hungyilee@ntu.edu.tw)}
}

\markboth{IEEE TRANSACTIONS ON AUDIO, SPEECH AND LANGUAGE PROCESSING, VOL. 34, 2026}{Chen \MakeLowercase{\textit{et al.}}: CodecFake+: Codec-Based Resynthesized Data as a Proxy for Detecting CodecFake Speech}

\maketitle

\begin{abstract}
With the rise of neural audio codecs, codec-based speech generation (CoSG) systems can produce highly realistic speech, posing new deepfake risks referred to as CodecFake. 
However, existing studies mainly focus on traditional synthesis methods, with limited exploration of CodecFake.
To address this gap, we introduce CodecFake+, the largest and most diverse dataset for codec-based deepfake detecton. 
It includes training data re-synthesized from 31 open-source neural codecs and evaluation data collected from 17 advanced CoSG models. 
We also propose a comprehensive taxonomy that categorizes codecs based on three key components: vector quantizer, auxiliary objectives, and decoder type. This taxonomy enables multi-level analysis to uncover key factors for detection. 
At the codec level, we validate the effectiveness of using codec re-synthesized speech (CoRS) for large-scale training. 
At the taxonomy level, we show that disentanglement objectives and frequency-domain decoders are primary determinants of codec forensic artifacts. 
Finally, we show that balancing training data across decoder types further enhances performance, even surpassing models trained solely on CoSG data. 
Overall, CodecFake+ provides a valuable foundation for advancing both general and fine-grained detection of codec-based deepfakes.
\end{abstract}


\begin{figure}[ht]
    \centering
    \includegraphics[width=1.0\columnwidth]{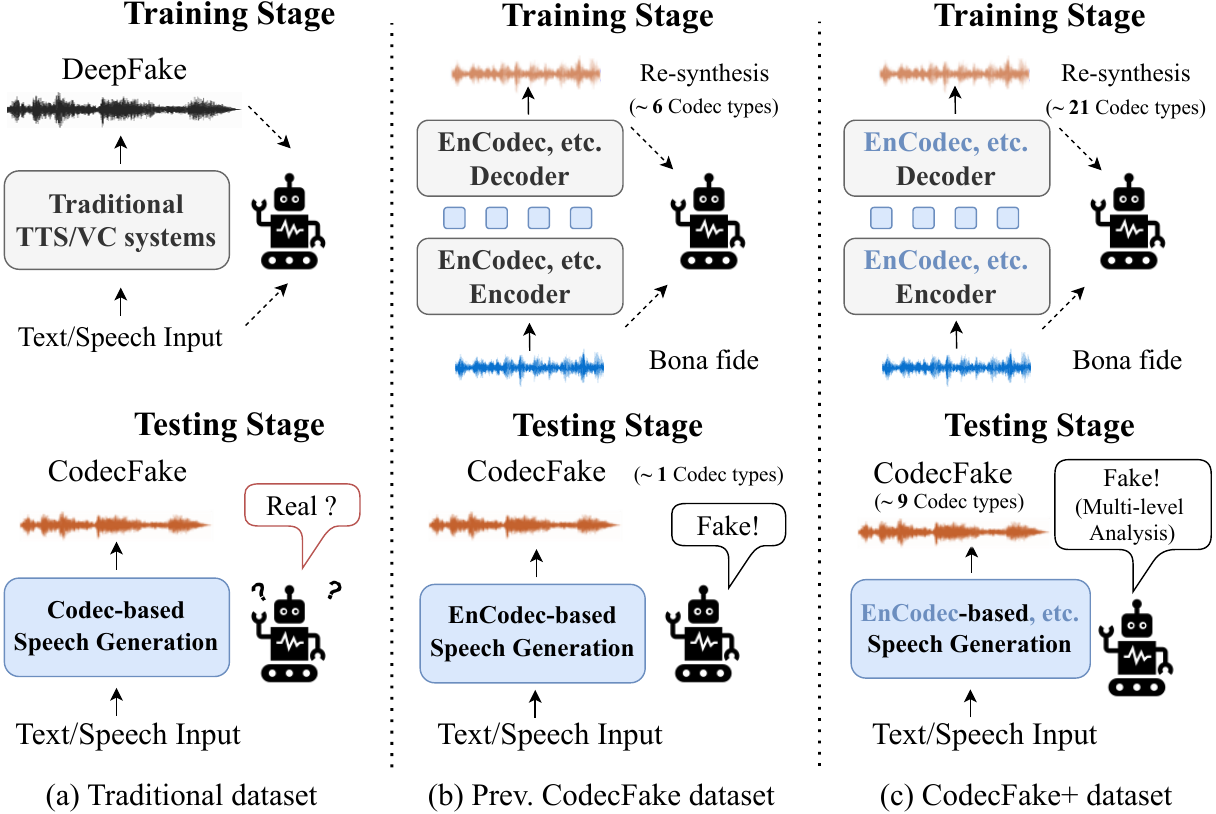}
    \vspace{-15pt}
    \caption{Illustration of different models detect CodecFake (i.e., speech generated by codec-based speech generation model) scenarios. 
    \textbf{(a) Traditional dataset:} Trained on conventional TTS/VC data, the model struggles with CodecFake detection.
    \textbf{(b) Previous CodecFake \cite{wu24p_interspeech} dataset:} A small-scale codec re-synthesis speech dataset for detecting EnCodec-based CodecFake.
    \textbf{(c) CodecFake+:} 
    A large-scale dataset enables better CodecFake detection and supports multi-level analysis.}
    \label{fig:trad_vs_omni}
    \vspace{-10pt}
\end{figure}

\section{Introduction}

\lettrine{A}{LTHOUGH} speech generation technology has advanced significantly in recent years,  
these techniques can be misused for malicious ``deepfake" attacks, leading to risks of misinformation, reputational damage, and manipulation of public opinion \cite{li2024audio, chintha2020recurrent, zhang2020deep, wu2023defender}.
To counteract the growing threat of deepfake attacks, initiatives such as the ASVspoof \cite{wu15e_interspeech, kinnunen17_interspeech, todisco2019asvspoof, nautsch2021asvspoof} and ADD \cite{yi2022add} challenges have been launched. 
These efforts have been instrumental in driving the development of advanced algorithms to defend against deepfake speech attacks \cite{wu15e_interspeech, kinnunen17_interspeech, todisco2019asvspoof, nautsch2021asvspoof, yi2022add, yi2023add, das2020attacker, wu2023defender}.
However, advances in data-driven neural audio codecs \cite{zeghidour2021soundstream, petermann2021harp_net, jiang2022end_tfnet, jiang2022cross_stfnet, jiang2023disentangled, betker2023better, borsos2023soundstorm, huang2023repcodec, lajszczak2024base_TTS, ai2024apcodec, gu2024esc, kim2024clam_mel-vae, pan2024promptcodec, guo2024addressing_pq-vae, zheng2024supercodec, an2024funaudiollm_S3Tokenizer, bie2024learning_sd-codec, zhou2024wmcodec, casanova2024low_lfsc, niu2024ndvq, zhang2023speechtokenizer, kumar2024high_DAC, yang2023hifi, defossez2022high, liu2024semanticodec, du2024funcodec, ju2024naturalspeech_FACodec, yang2024uniaudio_LLMCodec, ji2024wavtokenizer, ji2024language, ahn2024hilcodec, siuzdak2023vocos, guo2024socodec, siuzdak2024snac, ye2024codec, wu2023audiodec, xin2024bigcodec, ren2024fewer_ticodec, defossez2024moshi_mimi, langman2024spectral, yang2024simplespeech_sqcodec, li2024single, wu2024ts3, wu2024codec, mousavi2024dasb, wu2024codec_slt24, shi2024espnet, shi2024versa} have facilitated the development of codec-based speech generation (CoSG) systems \cite{wu2024towards}. 
Fake speech samples generated by these systems differ significantly from samples considered in earlier deepfake detection challenges such as ASVspoof and ADD. 
Whereas previous challenges primarily featured vocoder-related algorithms to convert synthesized acoustic features into speech, CoSG systems leverage neural audio codecs for modeling.
CoSG systems have advanced to the point where the voice of an unseen speaker can be reproduced using just a 3-second audio sample \cite{wang2023neural, wu2024towards, zhang2023speak, wang2023viola, yang2023uniaudio, du2023lauragpt, wang2024speechx, wu2024codec, wu2024codec_slt24}.
In addition, CoSG-generated speech is of increasingly high quality, often indistinguishable even to human listeners \cite{wang2023neural}.
We refer to the task of detecting fake speech samples generated by CoSG systems as \textbf{CodecFake} detection.

Our previous study \cite{wu24p_interspeech} showed that countermeasures (CMs) trained on traditional anti-spoofing datasets struggle with CodecFake detection (as shown in Figure \ref{fig:trad_vs_omni}.a), whereas training CMs on codec-resynthesized speech (CoRS) \cite{wu24p_interspeech} improves CodecFake detection (Figure \ref{fig:trad_vs_omni}.b). 
The performance gain stems from the shared decoder of CoSG and CoRS. 
Since both codecs reconstruct waveforms from codec features (discrete discrete codes or continuous embeddings) using the same process (codec decoder), they exhibit common artifacts. This consistency allows the model to better capture codec artifacts, thereby enhancing CodecFake detection performance \cite{wu24p_interspeech}.
Despite these gains, several key research questions remain:

\begin{enumerate}
    \item Which neural audio codec yields the most effective CoRS proxy for training CM against CodecFake attacks?
    \item What internal characteristics of a codec determine the suitability of its CoRS data?
    \item Can we improve CodecFake detection with more CoRS data selection?
\end{enumerate}

\begin{figure*}[t]
    \centering
    \includegraphics[width=2\columnwidth]{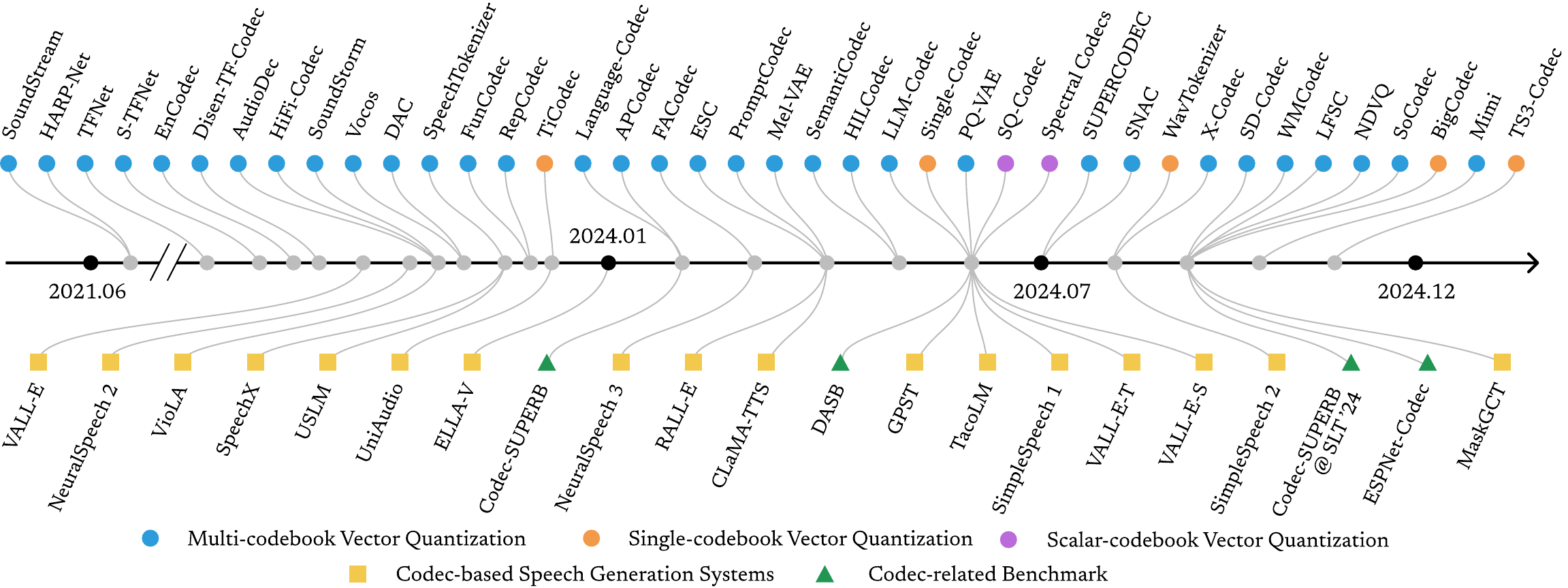}
    \vspace{-5pt}
    \caption{Timeline of current different types of neural audio codec models and codec-based speech generation models. Items above the horizontal line denote neural codec architectures, while those below the line represent derivative speech generation systems (CoSG) and established evaluation benchmarks. }
    \label{fig:timeline_codec_llm}
\end{figure*}

Since the release of our initial CodecFake database \cite{wu24p_interspeech}, numerous newer neural audio codecs and CoSG models have emerged, as shown in Figure~\ref{fig:timeline_codec_llm}.
Previous work \cite{muller22_interspeech} has highlighted how CM models may have poor generalization ability against unseen attacks. 
To allow future CM models to defend against increasingly diverse CoSG systems, we propose a large-scale dataset to systematically investigate how to leverage CoRS data for CM development.

Although leveraging CoRS data for CodecFake detection has been successful in previous studies \cite{wu24p_interspeech,lu24f_interspeech,xie2024codecfake},  the reasons behind its effectiveness remain unclear.
A taxonomy categorizing codecs by key characteristics would allow us to systematically identify which codec types or properties yield better detection performance and generalization. 
However, developing such a taxonomy is challenging, as no comprehensive overview of neural audio codecs existed prior to this work.

In this paper, we tackle the challenges in CodecFake detection by expanding our previously proposed CodecFake dataset~\cite{wu24p_interspeech}, and introducing a structured taxonomy for neural audio codecs. These steps enhance the scale and depth of CodecFake speech detection (as shown in Figure~\ref{fig:trad_vs_omni}.c)
We also provide a detailed multi-level analysis to understand how training data selection influences CodecFake detection performance. 
Our key contributions and findings as follow:
\begin{enumerate}

    \item We propose CodecFake+, which significantly scales the dataset compared to prior work \cite{wu24p_interspeech}, including nearly all publicly available codecs and CoSG models till February 2025. Specifically, we increase the number of distinct codecs for training data from 15 to 31 and expand the testing data from 3 to 17 CoSG models, representing a substantial extension. To our knowledge, CodecFake+ is the largest corpus for CodecFake detection to date.
    
    \item We propose a systematic taxonomy for neural audio codecs and CoSG, based on three attributes: vector quantizers, auxiliary objectives, and decoder types. The definition provides valuable insights for deeper analysis.

    \item Using the large-scale CodecFake+ dataset and well-defined neural audio codec taxonomy, we conducted a multi-level analysis for CodecFake detection, including codec-level, taxonomy-level, and database-level perspectives. 
    Instead of merely benchmarking performance, our study explores how various codec design choices affect detection, particularly in relation to vector-quantizer types, auxiliary objectives and decoder domains. This taxonomy offers a clear understanding of detection results and provides a solid framework for analyzing codec properties and selecting training data. 

\end{enumerate}

To support the research community in mitigating the risks posed by advanced CoSG technologies, we have publicly released all associated resources, 
including the demo page \footnote{Demo: \href{https://responsiblegenai.github.io/CodecFake-Plus-Dataset/}{https://responsiblegenai.github.io/CodecFake-Plus-Dataset/}}, 
implementation code \footnote{Code: \href{https://github.com/ResponsibleGenAI/CodecFake-Plus-Dataset}{https://github.com/ResponsibleGenAI/CodecFake-Plus-Dataset}}, 
and dataset \footnote{Data: \href{https://huggingface.co/datasets/CodecFake/CodecFake_Plus_Dataset}{https://huggingface.co/datasets/CodecFake/CodecFake\_Plus\_Dataset}.}. 
Resources are available after paper acceptance.

\section{Related Work}
Deepfake detection involves training CM models to differentiate between bona fide and fake audio, where dataset design is crucial. 
The ASVspoof challenges have progressively revealed vulnerabilities of automatic speaker verification (ASV) systems to deepfake speech: ASVspoof 2015 \cite{wu15e_interspeech} targeted text-to-speech (TTS) synthesis and voice conversion (VC) attacks; ASVspoof 2017 \cite{kinnunen17_interspeech} focused on replay attacks; and ASVspoof 2019–2021 \cite{nautsch2021asvspoof, liu2022asvspoof2021} addressed more realistic, real-world spoofing scenarios. 
CFAD \cite{ma2024cfad} and MLAAD \cite{10650962} expanded cross-lingual evaluations, while ASVspoof 5 \cite{Wang2024_ASVspoof5} introduced multilingual data, adversarial attacks, and a task-agnostic metric. With advances in generative modeling, new datasets have emerged targeting novel synthesis techniques, such as DFADD \cite{du2024dfadd} for diffusion-based deepfake and CodecFake \cite{wu24p_interspeech, xie2024codecfake} for CodecFake detection.

Our previous work~\cite{wu24p_interspeech} relied on a few neural codecs for CM training and three CoSG models for evaluation. 
To keep pace with fast‑evolving codecs and CoSG systems, we have scaled the dataset via the following key expansions:

\begin{enumerate} 
    \item \textit{Broader Neural Audio Codec Coverage:} For training data, our previous work \cite{wu24p_interspeech} leverages six neural codecs. Here, we expanded to 21 distinct neural codecs with different configurations, resulting in 31 different codec models for re-synthesizing speech, as shown in Table \ref{tab:codec_info}.

    \item \textit{Broader CoSG Coverage:} Previous studies primarily evaluated EnCodec-based CoSG systems. In contrast, this work broadens the evaluation by incorporating a diverse set of CoSG models based on 9 distinct codec types, resulting in a total of 17 CoSG systems (see Table \ref{tab:cosg_model}).
    
    \item \textit{Multi-Level Analysis:} Whereas our previous work primarily focused on codec-level analysis to examine the performance of models trained on a single CoRS set, this study conducts a comprehensive multi-level analysis:
    \begin{itemize}
        \item Codec-level analysis: Similar to our previous work, we independently train CMs on CoRS from each codec model, but on a much larger-scale CodecFake+ dataset.  
        Beyond that, we also analyze incorrectly classified samples, revealing that CoSG data with higher naturalness scores are more challenging to detect.

        \item Taxonomy-level analysis: We define a systematic neural audio codec taxonomy based on vector quantizers, auxiliary objectives, and decoders. 
        Taxonomy-based data selection not only boosts detection performance but also facilitates the analysis of key factors influencing generalization, such as the use of aligned vector quantization, disentanglement auxiliary objectives, and frequency-domain decoders in CoRS codecs. 

        \item Database-level analysis: 
        We propose several data selection strategies based on our taxonomy to improve CodecFake detection. 
        For example, balancing CoRS training data according to the decoder-type category yields the best performance. This shows our taxonomy improves understanding of CodecFake detection and provides effective guidance for data selection.
    \end{itemize}

\end{enumerate}

As shown in Table~\ref{tab:codecfake_dataset_comparison}, CodecFake+ offers the most comprehensive coverage of models and codec families (One codec type corresponds to a codec family and can produce multiple codec models based on different hyperparameter settings). 
And, CodecFake+ is the only dataset that supports multi-level analysis, providing deeper insights for researchers. 

\begin{figure}[t]
    \centering
        \includegraphics[width=0.9\columnwidth]{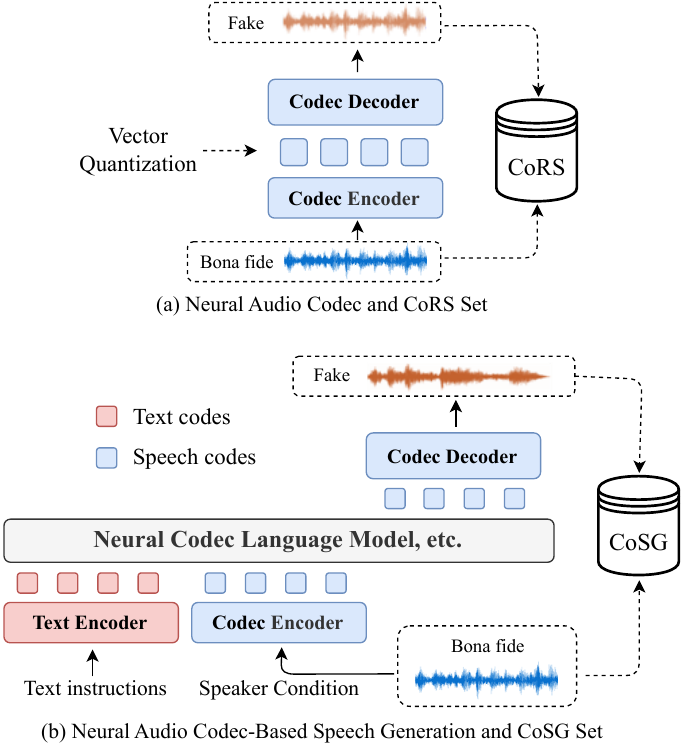}
    \vspace{-5pt}
    \caption{Comparison of neural audio codecs versus codec-based speech generation, and their CoRS (resynthesized) and CoSG (generated) datasets.}
    \label{fig:dataset_generation}
\end{figure}

\begin{table}[t!]
\centering
\fontsize{7}{10}\selectfont
\setlength\tabcolsep{2pt}
\renewcommand{\arraystretch}{1.2} 
\caption{Comparison with existing CodecFake datasets. 
Columns (a-b) represent the number of codec types used in the CoRS and CoSG sets. In Columns (c-d), we count codecs with different training configurations as distinct codec models.}
\label{tab:codecfake_dataset_comparison}
\begin{tabularx}{0.48\textwidth}{lccccccc}
\toprule
\multirow{2}{*}{\textbf{Dataset}} 
& \multicolumn{2}{c}{\#. of codec types}
& \multicolumn{2}{c}{\#. of codec models}
& \multicolumn{3}{c}{Multi-level analysis}  \\

\cmidrule(lr){2-3} \cmidrule(lr){4-5} \cmidrule(lr){6-8}
& (a) CoRS & (b) CoSG & (c) CoRS & (d) CoSG &  Codec  & Taxon. & Database  \\

\midrule
Wu et al. \cite{wu24p_interspeech} 
& 6   & 1  & 15 & 3   & {\color{Green} \cmark}  &  {\color{bsRed} \xmark} & {\color{Green} \cmark}   \\
Lu et al. \cite{lu24f_interspeech}
&  7  &  \textcolor{lightgray}{N/A} & 7 & \textcolor{lightgray}{N/A}  & {\color{bsRed} \xmark}  &  {\color{bsRed} \xmark} & {\color{Green} \cmark}    \\
Xie et al. \cite{xie2024codecfake}
&  7  &  2  & 14 & 3  & {\color{bsRed} \xmark}  &  {\color{bsRed} \xmark} & {\color{Green} \cmark}    \\
Xie et al. \cite{xie2024does} &  6  & 12  &  6  &  12  &  {\color{bsRed} \xmark}  &  {\color{bsRed} \xmark}  &  {\color{Green} \cmark}  \\
\midrule
CodecFake+ 
& 21 & 9 & 31 & 17  & {\color{Green} \cmark}  &  {\color{Green} \cmark} & {\color{Green} \cmark}  \\
\bottomrule
\end{tabularx}
\end{table}

\section{Conceptual Clarification: What is ``CodecFake?"}\label{sec:codecfake_definition}

Recent speech synthesis systems increasingly rely on neural audio codecs, where TTS or VC models generate codec representations (discrete or continuous) that are decoded into waveforms~\cite{mousavi2025discrete}. 
We refer to this paradigm as codec-based speech generation (CoSG). In this work, CodecFake denotes deepfake speech generated by such CoSG pipelines, namely audio that is not produced by a real human but synthesized through a codec-based speech generation system. Because its acoustic realization originates from a generative model rather than a real human, CodecFake is a category of deepfake audio.  

However, most CoSG systems are not publicly available, which makes it difficult to create training data generated in exactly the same way and to train models that specifically defend against CodecFake. 
A similar situation arises in traditional spoofing scenarios where unseen TTS or VC always appears. A common solution is copy‑synthesis through neural vocoder~\cite{wang2023spoofed}: converting speech to acoustic features and reconstructing it with neural vocoders to augment spoofing data for training. 
Just as a vocoder is the final stage in traditional synthesis, the final stage in many CoSG pipelines uses the same type of codec decoder as CoRS. As a result, speech reconstructed by a neural codec decoder can exhibit artifacts similar to those produced by codec‑based speech generation systems. Therefore, in CodecFake+ dataset, we treat CoRS as a practical \textbf{spoofing} proxy to train detectors to recognize these codec-induced artifacts. In this setup, all bona fide data remain unprocessed and identical to their originals. 

Please note that differing motivations, assumptions, and applications can lead to different definitions of codec‑resynthesized speech across benchmarks. For example, the ASVspoof5~\cite{Wang2024_ASVspoof5} evaluation set applies a neural codec (Encodec) to both bona fide and spoof classes for compression / augmentation purpose. In our study, codec‑resynthesized speech is treated as proxy spoof, following similar concept in vocoder based copy-synthesis. So that the model can learn artifacts introduced by the neural codec decoder, which is the final step in codec‑based speech generation. 
Xie \MakeLowercase{\textit{et al.}}~\cite{xie2024codecfake} 
adopt the same assumption in their CodecFake database, treating codec‑resynthesized speech as proxy spoof.
Definitions may vary by scenario and assumptions and should always be interpreted in context.

\section{Background of Neural Audio Codec and CoSG}
\label{sec:background}

We introduce the background of neural audio codecs and CoSG systems in this section 
\cite{wu2024towards, zeghidour2021soundstream, petermann2021harp_net, jiang2022end_tfnet, jiang2022cross_stfnet, jiang2023disentangled, betker2023better, borsos2023soundstorm, huang2023repcodec, lajszczak2024base_TTS, ai2024apcodec, gu2024esc, kim2024clam_mel-vae, pan2024promptcodec, guo2024addressing_pq-vae, zheng2024supercodec, an2024funaudiollm_S3Tokenizer, bie2024learning_sd-codec, zhou2024wmcodec, casanova2024low_lfsc, niu2024ndvq, zhang2023speechtokenizer, kumar2024high_DAC, yang2023hifi, defossez2022high, liu2024semanticodec, du2024funcodec, ju2024naturalspeech_FACodec, yang2024uniaudio_LLMCodec, ji2024wavtokenizer, ji2024language, ahn2024hilcodec, siuzdak2023vocos, guo2024socodec, siuzdak2024snac, ye2024codec, wu2023audiodec, xin2024bigcodec, ren2024fewer_ticodec, defossez2024moshi_mimi, langman2024spectral, yang2024simplespeech_sqcodec, li2024single, wu2024ts3, wu2024codec, mousavi2024dasb, wu2024codec_slt24, shi2024espnet,
song2024ella, wang2023neural,zhu2024generative_GPST, song2024tacolm, xin2024rall, wang2023viola, yang2024simplespeech_sqcodec, yang2024simplespeech2, wang2024maskgct, yang2023uniaudio, wang2024speechx, shen2024naturalspeech, eskimez2024e2ttsembarrassinglyeasy, kong2020hifi}.
Figure~\ref{fig:timeline_codec_llm} shows the evolution of neural audio codecs (top) and CoSG (bottom) models in CodecFake+, with Figure~\ref{fig:dataset_generation} illustrating their relationship.

\subsection{Neural Audio Codecs}

Figure~\ref{fig:timeline_codec_llm} shows the developments of neural audio codecs for audio compression, transmission, and speech–language modeling. 
As illustrated in Figure~\ref{fig:dataset_generation}.a, most codecs follow a three-stage framework: an encoder compresses audio, a vector quantizer (VQ) module converts features into discrete tokens, and a decoder reconstructs the waveform. 
The codec is vital for CoSG models, as they compress continuous speech to discrete tokens, which large language models and diffusion models can leverage. 
Several benchmarks, including Codec-SUPERB~\cite{wu2024codec,wu2024codec_slt24}, DASB~\cite{mousavi2024dasb}, and ESPnet-Codec~\cite{shi2024espnet}, cover at most six codec types; in contrast, CodecFake+ focuses on deepfake detection and extends codec coverage well beyond these (see Sec.~\ref{sec:CodecFake+_codec} and Table \ref{tab:codec_info}).
CoSG and CoRS often share a common architectural foundation: the neural audio codec. 
In CoSG, the model predicts codec features (tokens or embeddings), followed by a decoder to reconstruct back to speech. In CoRS, genuine speech is passed through the similar neural codec encode–decode path. For example, VALL-E uses EnCodec for decoding, we expect the speech resynthesized by Encodec showing similar forensic artifacts for detecting speech generated by VALL-E.

\subsection{Neural Codec-based Speech Generation Systems (CoSG)}

The bottom part of Figure~\ref{fig:timeline_codec_llm} shows the timeline of CoSG models used in CodecFake+.
Most systems rely on neural audio codec speech language modeling, with only a few using diffusion models (as indicated in Table~\ref{tab:cosg_model}). 
As shown in Figure \ref{fig:dataset_generation}.b, CoSG model encodes text information (e.g., instructions for expressiveness or target text content to be generated) into text codes and the speaker prompt audio (used to mimic the target speaker) into speech codes. 
These codes are processed in the neural codec language model (or diffusion models, etc.) to generate the target codec code sequence, which is then decoded by the codec decoder to produce the speech.
CoSG systems covered in this paper will be introduced in Section \ref{lab:CoSG_data} and Table~\ref{tab:cosg_model}.

\begin{figure}[t]
    \centering
\includegraphics[width=0.9\columnwidth]{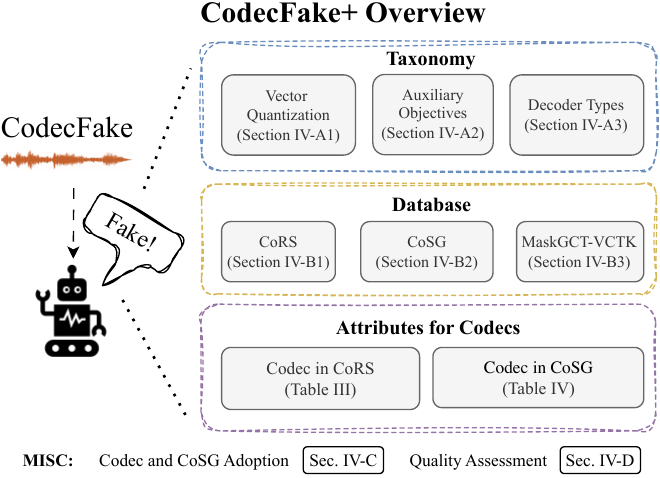}
    \vspace{-7pt}
    \caption{Organization of the CodecFake+ dataset. The taxonomy establishes the structural foundation for the multi-source database (comprising CoRS and CoSG subsets), while attribute categorization enables fine-grained analysis for CodecFake. Detailed descriptions are provided in Section~\ref{tab:sec_CodecFake+}.
    }
    \label{fig:codecfake+_overview}

\end{figure}

\section{CodecFake+ Database}
\label{tab:sec_CodecFake+}

Figure~\ref{fig:codecfake+_overview} provides an overview of CodecFake+ in three parts: Taxonomy (Section~\ref{lab:taxonomy}), Database (Section~\ref{sec:codecfake_omni}), and Categorization of Codecs (in CoRS Tables~\ref{tab:codec_info} and CoSG \ref{tab:cosg_model}). 
A detailed introduction of Codec and CoSG is in Section~\ref{sec:CodecFake+_codec_CoSG_detail}, followed by a quality assessment in Section~\ref{lab:quality_assessment}.

\subsection{Neural Audio Codec Taxonomy}
\label{lab:taxonomy}

The taxonomy for conventional TTS/VC models is well-defined (e.g., Table 1 in \cite{wang2020asvspoof}) in the existing studies, which consists of modular systems of acoustic models and vocoders. 
However, the rise of neural audio codecs has shifted speech generation toward codec-based architectures, where the codec decoder replaces the traditional vocoder. 
We use neural codec as a breakthrough point, as codec models are a common and crucial component in CoSG systems.
However, developing a taxonomy for neural audio codecs was challenging before this work due to the lack of a comprehensive overview, making the taxonomy presented in this paper novel and insightful. 
This taxonomy establishes a framework to link codec mechanisms with CodecFake artifacts. By focusing on core inductive biases including quantization, training objectives, and decoding domains, we clarify how architectural constraints shape synthetic traces. This allows for systematic categorization based on fundamental design choices, bypassing the need for an exhaustive analysis of every low-level parameter. 
Our Neural Audio Codec Taxonomy is based on three grouping categorizations: 
\textit{vector quantizer}, \textit{auxiliary objectives}, and \textit{decoder types}. 

\begin{table*}[t]
\centering
\caption{Summary of CodecFake+ Database. 
On the left, Table presents the bona fide data statistics across various subsets, while the right displays the statistics for spoof data.
In the first three rows, we present statistics for the re-synthesized speech from 31 codec models. In the CoSG Evaluation section, we provide statistics for CodecFake speech generated by 17 CoSG models. }
\label{tab:CodecFake+_Dataset}
\renewcommand{\arraystretch}{1.2} 
\fontsize{7}{8}\selectfont
\setlength\tabcolsep{8pt}

\begin{threeparttable}
\begin{tabular}{l ccc @{\hskip 0.5cm} l cc c c} 
\toprule
\multirow{2}{*}{\textbf{Subset Category}} & \multicolumn{3}{c}{\textbf{Bona fide Data}} & \multirow{2}{*}{\textbf{Subset Category}} & \multicolumn{4}{c}{\textbf{Spoof Data}} \\
\cmidrule(lr){2-4} \cmidrule(lr){6-9}
& Utter. & Total (h) & Min/Mean/Max (s) & & Utter. & Model & Total (h) & Min/Mean/Max (s) \\
\midrule

\rowcolor{gray!10} \multicolumn{9}{l}{\textit{Codec Re-synthesized (CoRS) Part}} \\
CoRS Train       & 42,965 & 40.10 & 1.22 / 3.36 / 16.56 & CoRS Train       & 42,965 $\times$ 31 & 31 & 1261.61 & 1.20 / 3.41 / 17.28 \\
CoRS Validation  & 735    & 0.69  & 1.62 / 3.40 / 12.15 & CoRS Validation  & 735 $\times$ 31    & 31 & 21.77   & 1.61 / 3.44 / 12.84 \\
CoRS Evaluation  & 755    & 0.76  & 1.49 / 3.63 / 12.68 & CoRS Evaluation  & 755 $\times$ 31    & 31 & 23.86   & 1.48 / 3.67 / 13.32 \\
\midrule

\rowcolor{gray!10} \multicolumn{9}{l}{\textit{CodecFake Evaluation Dataset generated by CoSG }} \\
CoSG Evaluation  & 850    & 1.69  & 1.27 / 6.55 / 28.58 & CoSG Evaluation  & 931                & 17 & 1.32    & 0.85 / 5.51 / 16.63 \\
\midrule

\rowcolor{gray!10} \multicolumn{9}{l}{\textit{MaskGCT-VCTK CoSG Baseline (Proposed Comparison Group)}} \\
MaskGCT-VCTK Tr.\tnote{1} & 42,792 & 40.10 & 1.22 / 3.36 / 16.56 & MaskGCT-VCTK Tr.\tnote{1} & 42,792 & 1 & 42.23 & 0.56 / 3.55 / 18.76 \\
MaskGCT-VCTK Val.         & 735    & 0.69  & 1.62 / 3.40 / 12.15 & MaskGCT-VCTK Val.         & 735    & 1 & 0.75  & 0.76 / 3.71 / 13.12 \\
MaskGCT-VCTK Eva.         & 755    & 0.76  & 1.49 / 3.63 / 12.68 & MaskGCT-VCTK Eva.         & 755    & 1 & 0.81  & 0.84 / 3.87 / 16.30 \\
MaskGCT-E2\tnote{2}       & 576    & 1.04  & 4.00 / 6.50 / 10.00 & MaskGCT-E2\tnote{2}       & 576    & 1 & 1.18  & 2.58 / 7.38 / 28.58 \\
\bottomrule
\addlinespace[1.5pt]
\end{tabular}
\begin{tablenotes}
    \item [1] The speaker p315 lacks transcriptions from the officially released VCTK, causing slight sample discrepancies between MaskGCT-VCTK Train and CoRS Train. 
    \item [2] MaskGCT-E2, sourced from E2 TTS test set, is also included in the row `CoSG Evaluation.'
\end{tablenotes}
\end{threeparttable}
\end{table*}

\subsubsection{Vector quantization}
Vector quantization transforms audio frames into discrete tokens, classified into three types:
\begin{itemize}
    \item \textbf{Multi-codebook vector quantization (Mvq)} compresses audio frames using sequential or parallel quantizers. Examples include EnCodec \cite{defossez2022high} with residual vector quantization, ESC \cite{gu2024esc} with cross-scale residual quantization and hierarchical transformers, and HiFi-Codec \cite{yang2023hifi} with group residual vector quantization with fewer codebooks.

    \item \textbf{Single-codebook vector quantization (Svq)} uses a single codebook for quantization. Examples include TiCodec \cite{ren2024fewer_ticodec} and Single-Codec \cite{li2024single}, which encode global information into an utterance-level vector and time-varying information into a single stream of codes. ~\cite{wu2024ts3,xin2024bigcodec} push extremely low-bitrate high-quality Svq codecs. 

    \item \textbf{Scalar vector quantization (Scq)} simplifies audio representation by mapping speech signals into a structured, low-dimensional latent space. Unlike traditional vector quantization, Scq effectively alleviates the codebook collapse and improves the codebook utilization. 
    SQ-Codec \cite{yang2024simplespeech_sqcodec} achieves this by creating a scalar latent space, while Spectral Codecs \cite{langman2024spectral} utilize finite scalar quantization to encode mel-spectrogram features into a flat codebook.
\end{itemize}

\subsubsection{Auxiliary objectives}
Auxiliary objectives refer to the auxiliary loss for enhancing generalization, capabilities, or desired properties. We have three types of auxiliary objectives:
\begin{itemize}
    \item \textbf{Disentanglement (Disent)} in neural audio codecs separates speech attributes into distinct representations. For instance, TiCodec \cite{liu2024semanticodec} separates static and dynamic features to reduce token rates. FACodec \cite{ju2024naturalspeech_FACodec} factorizes waveforms into content, prosody, and timbre. SoCodec \cite{guo2024socodec} employs two codec encoders to model frame-wise semantic information and global information.

    \item \textbf{Semantic distillation (Sem)} refers to techniques that enhance codec codes with semantic information. 
    Methods include pre-trained models for semantic guidance (SpeechTokenizer \cite{zhang2023speechtokenizer}, X-Codec \cite{ye2024codec}, Mimi Codec \cite{defossez2024moshi_mimi}), dual-encoder compression (SemantiCodec \cite{liu2024semanticodec}), and tokenization for LLM compatibility (LLM-Codec \cite{yang2024uniaudio_LLMCodec}). 
    
    \item \textbf{None} refer to no auxiliary objectives.
\end{itemize}

\subsubsection{Decoder type}
The decoder in audio codecs reconstructs waveforms (usually 16k or 24k Hz) from discrete units (usually 50 Hz), with upsampling as a key operation. This can be achieved using two methods based on processing domain:
\begin{itemize}
    \item \textbf{Time-domain decoder (Time)} mainly utilizes transposed convolution for upsampling the codec representation to raw audio waveforms, such as EnCodec \cite{defossez2022high}, SpeechTokenizer \cite{zhang2023speechtokenizer}, DAC \cite{kumar2024high_DAC}, HiFiCodec \cite{yang2023hifi}. 
    
    \item \textbf{Frequency-domain decoder (Freq)} employs the Inverse Short-Time Fourier Transform on the decoded features from the decoder (decoded features have a similar frame rate with the codec tokens) for upsampling, in frequency-domain codecs such as WavTokenizer \cite{ji2024wavtokenizer}, LanguageCodec \cite{ji2024language}, Vocos \cite{siuzdak2023vocos}, and Spectral Codecs \cite{langman2024spectral}. 
\end{itemize}

The taxonomy provides a structured framework for understanding existing codecs and guiding the development of CodecFake detection models. 
Based on this, we categorize various neural audio codecs and CoSG systems within CodecFake+ as shown in Table \ref{tab:codec_info} and Table \ref{tab:cosg_model}.
This taxonomy can also be applied to cover other neural audio codecs or CoSG systems beyond CodecFake+ database in future research.
Additionally, this taxonomy offers insights for future CodecFake research, particularly in source tracing, domain adaptation, and domain generalization for detecting previously unseen CoSG.

\subsection{CodecFake+ Dataset} \label{sec:codecfake_omni}

For training data, we use publicly available pre-trained neural audio codec models to re-synthesize speech (named CoRS) following \cite{wu24p_interspeech}, and we also created an evaluation set in this way for reference. 
For testing data, we collect web-sourced data from unreleased CoSG models. The relationship between CoRS and CoSG is indicated in Figure~\ref{fig:dataset_generation}.
In addition, although the web-sourced CoSG data is suitable for testing but limited and insufficient for training, we further explore CM training on CoSG by creating  
an additional train/validation/evaluation set using the open-source CoSG model MaskGCT \cite{wang2024maskgct}. 
The dataset statistics are shown in Table~\ref{tab:CodecFake+_Dataset}

\subsubsection{CoRS Train/Validation/Evaluation Sets}\label{lab:cf-omni_training_data}

CoRS in CodecFake+ are generated via codec re-synthesis (Figure~\ref{fig:dataset_generation}.a). 
It contains three subsets: training, validation, and evaluation.
Following CodecFake \cite{wu24p_interspeech}, we generate CodecFake+ using VCTK corpus \cite{yamagishi2019cstr}. 
The 110 speakers are split into three disjoint sets: two speakers (p226 and p229) for validation, two speakers (p227 and p228) for Eval CoRS, and the remaining speakers for training. 
Here, ``pXXX" refers to the speaker identifiers in the VCTK dataset.  
We use 31 open-source codecs to re-synthesize speech to construct spoofing CoRS data, while the original speech from VCTK is treated as bona fide.
All subsets include both bona fide and spoofed speech.
As shown in Figure~\ref{fig:dataset_generation}.a, pre-trained codec models encode, quantize, and reconstruct bona fide speech. 
Each set of re-synthesized audio is paired with its bona fide input to get a CoRS subset. 
Although the initial sampling rate of the re-synthesized data follows the original setup of the pre-trained codecs, ranging from 16 kHz to 44.1 kHz, all data are resampled to 16 kHz for consistency in CM training and evaluation.

\subsubsection{CoSG Evaluation Set (Figure~\ref{fig:dataset_generation}.b)}
\label{lab:codecfake_omni_eval_cosg}

Most CoSG models are unavailable for privacy reasons, so we collected data from their official demo websites up to February 2025.
This better simulates real-world conditions where deepfake generation systems are unseen.  
For web-sourced downloaded data, speaker prompts spoken by target speakers are treated as bona fide data, and generated samples from CoSG models are labeled as spoofs. 
Note that because speaker prompts are used to provide the target speaker’s characteristics, the bona fide and spoofed speech are not text-aligned. More information about each CoSG model can be found in the corresponding papers or links listed in Section~\ref{lab:CoSG_data}
Table~\ref{tab:cosg_model} provides details on test set. 
For spoofed samples produced via speech continuation or editing, we rigorously trimmed each sample to retain only the generated regions.
All audio was standardized to 16kHz without loudness normalization to preserve original speech synthesis artifacts. While multilingual deepfake detection is important, CodecFake+ utilizes an controlled English-only dataset to isolate forensic signatures tied to neural audio codec artifacts and their structural diversity. 

\subsubsection{MaskGCT-VCTK Train/Validation/Evaluation Sets and MaskGCT-E2 Evaluation Set}
\label{lab:maskgct-cf_cosg_baseline}

While the aforementioned CoRS sets cover a broad range of codecs, they were re-synthesized using pre-trained codec models, as opposed to being generated by CoSG.  
To evaluate CMs trained on CoSG data as baselines, we used MaskGCT, an open-source CoSG model, to create another subset for comparison.

Few CoSG models are open-sourced. Given two key considerations, namely the availability of open-source CoSG models and their ability to generate high-quality speech \cite{jia2025ditar}, we ultimately selected MaskGCT \cite{wang2024maskgct}. 
To simulate a realistic scenario where evaluation sets are usually in the unseen condition, we use two different source data (with different types of speaker prompts and transcriptions): MaskGCT-VCTK uses VCTK \cite{yamagishi2019cstr} as the source dataset for generating training and validation sets, resulting in MaskGCT-VCTK Train/Validation/Evaluation Sets. 
 MaskGCT-E2 Evaluation Set uses the E2 TTS test set \cite{eskimez2024e2ttsembarrassinglyeasy} as the source dataset for generating the CoSG evaluation set. The original data from these datasets are treated as bona fide for their respective sets. 
During audio synthesis, utterances are sorted alphabetically per speaker, using the current waveform as the audio prompt and the next transcript as the text prompt.
For the last utterance, it loops back to the first transcript.

\begin{table}[t!]
\centering
\renewcommand{\arraystretch}{1.2} 
\fontsize{7}{8}\selectfont
\setlength\tabcolsep{1.8pt}
\caption{Neural audio codecs in CoRS dataset of CodecFake+. Arch. denotes the neural network architecture (Encoder and Decoder). QUA, AUX, and DEC represent vector quantization, auxiliary objectives, and decoder types. }
\label{tab:codec_info}
\begin{tabularx}{0.48\textwidth}{cccc ccc}
    \toprule
    \multicolumn{3}{c}{\textbf{Codec-level}} &  & \multicolumn{3}{c}{\textbf{Taxonomy-level}} \\
    \cmidrule(r){2-4} \cmidrule(l){5-7}
    \textbf{ID} & \textbf{Codec in CoRS} & \textbf{Config} & \textbf{Arch.$^*$} & \textbf{QUA} &  \textbf{AUX} & \textbf{DEC}  \\
    \midrule

    A & SpeechTokenizer \cite{zhang2023speechtokenizer} & hubert\_avg & CR / C & Mvq &  Sem & Time  \\

    \rowcolor{gray!10} B & DAC \cite{kumar2024high_DAC} & DAC24k & C / C & Mvq & None & Time   \\

    C & HiFi-Codec  \cite{yang2023hifi}     & { \fontsize{6pt}{8}\selectfont  16k\_320d\_large\_universal} & CR / CR & Mvq & None & Time  \\

    \rowcolor{gray!10} D1 &  & { \fontsize{6pt}{8}\selectfont  6b24k} & CR / C  & Mvq & None & Time  \\
    \rowcolor{gray!10} D2 & \multirow{-2}{*}{EnCodec \cite{defossez2022high}} & { \fontsize{6pt}{8}\selectfont  24b24k} & CR / C  & Mvq & None & Time   \\
    
    E1 &  \multirow{3}{*}{SemantiCodec \cite{liu2024semanticodec}} & 1.40kbps\_16k & CT / T & Mvq &  Sem  &  Freq  \\
    E2 & & 0.70kbps\_16k & CT / T & Mvq &  Sem  &  Freq  \\
    E3 & & 0.35kbps\_16k & CT / T & Mvq &  Sem  &  Freq  \\

    \rowcolor{gray!10} F1 &  & { \fontsize{6pt}{8}\selectfont  en-libritts-16k-gr1nq32ds320} & CR / CR & Mvq & None & Freq \\
    \rowcolor{gray!10} F2 & \multirow{-2}{*}{FunCodec \cite{du2024funcodec}} & { \fontsize{6pt}{8}\selectfont  zh\_en-general-16k-nq32ds320} & CR / CR & Mvq & None & Time \\
    
    G & FACodec \cite{ju2024naturalspeech_FACodec} & { \fontsize{6pt}{8}\selectfont  encodec-decoder-v2\_16k} & CR / CR & Mvq & Disent & Time \\
    
    \rowcolor{gray!10} H & LLM-Codec \cite{yang2024uniaudio_LLMCodec} & llm\_codec & CT / CT & Mvq & Sem & Time \\
    
    I1 & \multirow{2}{*}{WavTokenizer \cite{ji2024wavtokenizer}} & small\_320\_24k\_4096 & CT / CT & Svq & None & Freq \\
    
    I2 & & medium\_320\_24k\_4096 & CT / CT & Svq &  None & Freq \\
    
    \rowcolor{gray!10} J & Language-Codec \cite{ji2024language} & language-codec & CR / C & Mvq &  None & Freq \\
    
    K & HILCodec \cite{ahn2024hilcodec} & hilcodec\_speech & C / C & Mvq &  None  &  Time \\
    
    \rowcolor{gray!10} L1 &  & vocos\_encodec\_6 & C / C & Mvq &  None &  Freq \\
    \rowcolor{gray!10} L2 & \multirow{-2}{*}{Vocos \cite{siuzdak2023vocos}} & vocos\_encodec\_12 & C / C & Mvq &  None &  Freq \\
    
    M & SoCodec \cite{guo2024socodec} & {\fontsize{6pt}{8}\selectfont 16384x4\_120ms\_16khz\_chinese} & C / C &  Mvq  &  Disent  &  Freq  \\

    \rowcolor{gray!10} P1 &  & 24khz & CR / C & Mvq &  None  & Time \\
    \rowcolor{gray!10} P2 & \multirow{-2}{*}{SNAC \cite{siuzdak2024snac}} & 44khz & CR / C & Mvq &  None  &  Time \\
    
    Q1 & \multirow{2}{*}{X-Codec \cite{ye2024codec}} & hubert\_librispeech & C / C & Mvq & Sem & Time  \\
    Q2 & & hubert\_general\_audio & C / C & Mvq & Sem & Time  \\
    
    \rowcolor{gray!10} R & AudioDec \cite{wu2023audiodec} & audiodec\_24k\_320d & C / C &   Mvq & None & Freq \\
    
    S & BigCodec \cite{xin2024bigcodec} & BigCodec & CR / CR & Svq & None & Time \\

    \rowcolor{gray!10} T1 &  & ticodec\_1g1r & CR / CR & Svq & Disent & Time \\
    \rowcolor{gray!10} T2 &  & ticodec\_1g2r & CR / CR & Mvq &  Disent & Time  \\
    \rowcolor{gray!10} T3 & \multirow{-3}{*}{TiCodec \cite{ren2024fewer_ticodec}} & ticodec\_1g4r & CR / CR & Mvq &  Disent & Time  \\
    
    U & Mimi \cite{defossez2024moshi_mimi} & mimi-codec & CT / CT & Mvq & Sem & Time  \\
    
    \rowcolor{gray!10} V & Spectral Codecs \cite{langman2024spectral} & spectral codecs & C / C & Scq &  None  & Freq  \\

    W & SQ-Codec \cite{yang2024simplespeech_sqcodec} & sqcodec50dim9  & C / C & Scq &  None & Time  \\
    \bottomrule
    \multicolumn{7}{l}{$^*$ Note: C, T, and R denote CNN, Transformer, and RNN, respectively. }
\end{tabularx}
\end{table}

\subsection{Neural Audio Codec and CoSG in CodecFake+}\label{sec:CodecFake+_codec_CoSG_detail}
\subsubsection{Neural codec in CodecFake}\label{sec:CodecFake+_codec}
We include re‑synthesized speech from 31 models across 21 codec (Table~\ref{tab:codec_info} families with varied configurations, forming Train and Eval CoRS sets. 

\textbf{EnCodec (D)} \cite{defossez2022high} is a pioneer model with a typical design with separate encoder, quantizer, and decoder modules, enhanced by LSTM and Transformer-based sequence modeling. 

\textbf{SpeechTokenizer (A)} \cite{zhang2023speechtokenizer} uses HuBERT to guide semantic distillation in a Residual Vector Quantizer, where the first layer encodes semantics and later layers capture acoustic details. 

\textbf{Descript Audio Codec (B)} \cite{kumar2024high_DAC}, or DAC, is a high-fidelity neural codec that enhances vector quantization and loss functions for strong compression, supporting various audio. 

\textbf{HiFi-Codec (C)} \cite{yang2023hifi} employs Group Residual Vector Quantization for hybrid compression, enabling high-quality, efficient audio synthesis with fewer codebooks. 

\textbf{SemantiCodec (E)} \cite{liu2024semanticodec} uses a dual encoder: a semantic encoder based on AudioMAE \cite{NEURIPS2022_b89d5e20} with k-means discretization, and an acoustic encoder capturing finer-grained details.

\textbf{FunCodec (F)} \cite{du2024funcodec} includes two codec models in the time and frequency domains, with the frequency-domain model achieving comparable performance using fewer parameters.

\textbf{FACodec (G)} \cite{ju2024naturalspeech_FACodec} achieves high-quality reconstruction by factorizing speech into prosody, timbre, and acoustic subspaces, using gradient reversal for disentanglement.

\textbf{LLM-Codec (H)} \cite{yang2024uniaudio_LLMCodec} maps audio into Large Language Model (LLM) token space, enabling LLMs to process audio as a ``foreign language" with examples. LLMs can perform audio tasks without parameter updates, using a few prompts. 

\textbf{WavTokenizer (I)} \cite{ji2024wavtokenizer} is a frequency-domain codec using a single quantizer for compression, reducing 24kHz audio to 40–75 tokens per second while preserving naturalness. 

\textbf{Language-Codec (J)} \cite{ji2024language} enhances speech-language models using Masked Channel Residual Vector Quantization, balancing channel information for improved audio restoration. 

\textbf{HILCodec (K)} \cite{ahn2024hilcodec} is a high-fidelity, lightweight streaming codec that uses variance constraints for stability and a Multi-Filter Bank Discriminator to minimize aliasing and distortion.

\textbf{Vocos (L)} \cite{siuzdak2023vocos} is a vocoder that generates Fourier spectral coefficients, bridging time-domain GANs and frequency-based representations. Delivers high-quality audio with an order-of-magnitude efficiency gain. Combined with an EnCodec encoder and quantizer, it forms a neural audio codec. 

\textbf{SoCodec (M)} \cite{guo2024socodec} applies Ordered Product Quantization for semantic compression, separating core and fine-grained features across streams to enhance stability and accuracy. 

\textbf{SNAC (P)} \cite{siuzdak2024snac} enhances Residual Vector Quantization with multi-scale quantizers at varying resolutions, achieving higher audio quality at lower bitrates.

\textbf{X-Codec (Q)} \cite{ye2024codec} utilizes a pre-trained semantic encoder and reconstruction loss to enrich token semantics. Its unified tokenization framework operates independently of LLM. 

\textbf{AudioDec (R)} \cite{wu2023audiodec} extends EnCodec with group convolution for real-time streaming and uses HiFi-GAN \cite{kong2020hifi} to generate high-fidelity 48 kHz audio. 

\textbf{BigCodec (S)} \cite{xin2024bigcodec} is a large-scale codec optimized for \~1 kbps bitrates using a single codebook to achieve high-fidelity speech reconstruction at ultra-low bitrates.

\textbf{TiCodec (T)} \cite{ren2024fewer_ticodec} produces fixed-length, time-invariant tokens with frame reduction via a specialized extractor, using a consistency loss for stable and effective representations. 

\textbf{Mimi (U)} \cite{defossez2024moshi_mimi} is a Transformer-based streaming codec for Moshi dialogue framework, modeling semantic and acoustic features with auxiliary losses to improve low-bitrate quality.
 
\textbf{Spectral Codecs (V)} \cite{langman2024spectral} compress mel-spectrograms for frequency-domain reconstruction by scalar-quantizing disjoint mel bands and concatenating embeddings for decoding.

\textbf{SQ-Codec (W)} \cite{yang2024simplespeech_sqcodec} employs scalar quantization to encode speech into a compact latent space. It normalizes features via a hyperbolic tangent function, then rescales for discretization.

\begin{table}[t!]
\centering
\fontsize{6.5}{9}\selectfont
\setlength\tabcolsep{1.5pt}
\renewcommand{\arraystretch}{1.2} 
\caption{Statistics of the CoSG in CodecFake+ dataset. The column `Number' presents in \#. of Bona fide / \#. of Spoof utterances.} 
\label{tab:cosg_model}
\begin{tabularx}{0.48\textwidth}{lcccccc}
\toprule
    \multicolumn{2}{c}{\textbf{Codec-level}} & \multicolumn{3}{c}{\textbf{Taxonomy-level}}  & \textbf{Number} &  \multirow{2}[0]{*}{\textbf{Source}} \\
    \cmidrule(r){1-2} \cmidrule(lr){3-5} \cmidrule(lr){6-6}
\textbf{CoSG Model} &  \textbf{Codec in CoSG} & \textbf{QUA}  & \textbf{AUX} &  \textbf{DEC} & \textbf{Bona. / Spoof} &  \\
\midrule

\rowcolor{gray!10}[\tabcolsep][\dimexpr\tabcolsep+0.5pt\relax] ELLA-V* \cite{song2024ella} & EnCodec \cite{defossez2022high}  & Mvq & None & Time & 8 / 8  & Demo \\
\rowcolor{gray!10}[\tabcolsep][\dimexpr\tabcolsep+0.5pt\relax] VALL-E* \cite{wang2023neural} & EnCodec \cite{defossez2022high} & Mvq & None & Time & 53 / 57 & Demo \\
\rowcolor{gray!10}[\tabcolsep][\dimexpr\tabcolsep+0.5pt\relax] GPST* \cite{zhu2024generative_GPST} & EnCodec \cite{defossez2022high} & Mvq & None & Time & 10 / 30 & Demo\\ 
\rowcolor{gray!10}[\tabcolsep][\dimexpr\tabcolsep+0.5pt\relax] UniAudio* \cite{yang2023uniaudio} & EnCodec \cite{defossez2022high} & Mvq & None & Time & 11 / 11 & Demo \\ 
\rowcolor{gray!10}[\tabcolsep][\dimexpr\tabcolsep+0.5pt\relax] TacoLM* \cite{song2024tacolm} & EnCodec \cite{defossez2022high} & Mvq & None & Time & 8 / 8 & Demo \\ 
\rowcolor{gray!10}[\tabcolsep][\dimexpr\tabcolsep+0.5pt\relax] SpeechX* \cite{wang2024speechx} & EnCodec \cite{defossez2022high} & Mvq & None & Time & 8 / 8 & Demo \\
\rowcolor{gray!10}[\tabcolsep][\dimexpr\tabcolsep+0.5pt\relax] RALL-E* \cite{xin2024rall} & EnCodec \cite{defossez2022high} & Mvq & None & Time & 5 / 5 & Demo \\ 
\rowcolor{gray!10}[\tabcolsep][\dimexpr\tabcolsep+0.5pt\relax] VioLA* \cite{wang2023viola} & EnCodec \cite{defossez2022high} & Mvq & None & Time & 7 / 7 & Demo \\

CLaM-TTS* \cite{kim2024clam_mel-vae} & Mel-VAE \cite{kim2024clam_mel-vae} & Mvq & None & Freq & 41 / 78  & Demo \\

\rowcolor{gray!10}[\tabcolsep][\dimexpr\tabcolsep+0.5pt\relax] NaturalSpeech 2 \cite{shen2024naturalspeech} & SoundStream \cite{zeghidour2021soundstream} & Mvq & None & Time & 23 / 23 & Demo \\

NaturalSpeech 3 \cite{ju2024naturalspeech_FACodec} & FA-Codec \cite{ju2024naturalspeech_FACodec} & Mvq & Disent & Time &  32 / 32 & Demo \\
 
\rowcolor{gray!10}[\tabcolsep][\dimexpr\tabcolsep+0.5pt\relax] USLM* \cite{zhang2023speechtokenizer} & SpeechTokenizer \cite{zhang2023speechtokenizer} & Mvq & Sem & Time & 5 / 5 & Demo \\

SimpleSpeech 1\cite{yang2024simplespeech_sqcodec} & SQ-Codec \cite{yang2024simplespeech_sqcodec} & Scq & None & Time & 34 / 32 & Demo \\
SimpleSpeech 2 \cite{yang2024simplespeech2} & SQ-Codec \cite{yang2024simplespeech_sqcodec} & Scq & None & Time & 31 / 31 & Demo \\

\rowcolor{gray!10}[\tabcolsep][\dimexpr\tabcolsep+0.5pt\relax] VALL-E-T* \cite{ren2024fewer_ticodec} & TiCodec \cite{ren2024fewer_ticodec} & Svq & Disent & Time & 6 / 6 & Demo \\

VALL-E-S* \cite{li2024single} & Single-Codec \cite{li2024single} & Svq & Disent & Freq & 8 / 14 & Demo \\

\rowcolor{gray!10}[\tabcolsep][\dimexpr\tabcolsep+0.5pt\relax] MaskGCT-E2 \cite{wang2024maskgct} & DAC-Vocos \cite{wang2024maskgct} & Mvq & None & Freq & 576 / 576 & E2 TTS \cite{eskimez2024e2ttsembarrassinglyeasy} \\
\bottomrule
\addlinespace[1.5pt]
\multicolumn{7}{l}{* denotes language modeling based CoSG systems.}
\end{tabularx}
\end{table}

\subsubsection{CoSG model in CodecFake+}
\label{lab:CoSG_data}

To assess CodecFake speech detection, we gathered data from 17 models (Table~\ref{tab:cosg_model}). 

\textbf{ELLA-V}\footnote{\href{https://ereboas.github.io/ELLAV/}{https://ereboas.github.io/ELLAV/}} \cite{song2024ella} is a zero-shot TTS model based on EnCodec. 
It achieves phoneme-level control by interleaving phoneme and acoustic tokens and preordering phonemes.

\textbf{VALL-E} \footnote{\href{https://www.microsoft.com/en-us/research/project/vall-e-x/vall-e/}{https://www.microsoft.com/en-us/research/project/vall-e-x/vall-e/}} \cite{wang2023neural} treats TTS as conditional language modeling via EnCodec and an autoregressive model, producing expressive speech using a three-second prompt.

\textbf{GPST}\footnote{\href{https://youngsheen.github.io/GPST/demo/}{https://youngsheen.github.io/GPST/demo/}} \cite{zhu2024generative_GPST} is a hierarchical transformer that quantizes audio using self-supervised k-means and EnCodec, enabling unified one-stage and high-resolution speech generation. 

\textbf{UniAudio}\footnote{\href{https://dongchaoyang.top/UniAudio_demo/}{https://dongchaoyang.top/UniAudio\_demo/}} \cite{yang2023uniaudio} tokenizes inputs and outputs into a single sequence and employs a multi‑scale Transformer to process long EnCodec token streams.

\textbf{TacoLM}\footnote{\href{https://ereboas.github.io/TacoLM/}{https://ereboas.github.io/TacoLM/}} \cite{song2024tacolm} is a model with a gated attention mechanism for efficiency, improved synthesis accuracy, and reduced model size, using pre-trained EnCodec as its audio tokenizer. 

\textbf{SpeechX}\footnote{\href{https://www.microsoft.com/en-us/research/project/speechx/}{https://www.microsoft.com/en-us/research/project/speechx/}} \cite{wang2024speechx} is a unified CoSG model integrating EnCodec-based neural codec language modeling and multi-task learning with task-specific prompting. 

\textbf{RALL-E}\footnote{\href{https://ralle-demo.github.io/RALL-E/}{https://ralle-demo.github.io/RALL-E/}} \cite{xin2024rall}, an EnCodec-based extension of VALL-E, introduces chain-of-thought prompting by predicting pitch and duration. Then, refining attention with duration prompts. 

\textbf{CLaM-TTS}\footnote{\href{https://clam-tts.github.io/}{https://clam-tts.github.io/}} \cite{kim2024clam_mel-vae} introduces a mel-based codec with high compression, enabling CoSG models to generate multiple token stacks simultaneously, removing cascaded modeling.

\textbf{VioLA}\footnote{\href{https://violademo.github.io/}{https://violademo.github.io/}} \cite{wang2023viola} is a multilingual and multimodal transformer built on VALL‑E. 
With EnCodec tokenization, speech is treated as text, enabling modeling via a decoder‑only model. 
 
\textbf{NeuralSpeech 2} \footnote{\href{https://speechresearch.github.io/naturalspeech2/}{https://speechresearch.github.io/naturalspeech2/}} \cite{shen2024naturalspeech} is a CoSG model leveraging latent diffusion and SoundStream with residual VQ for zero-shot, multilingual, and robust speech synthesis. 
 
\textbf{NeuralSpeech 3}\footnote{\href{https://speechresearch.github.io/naturalspeech3/}{https://speechresearch.github.io/naturalspeech3/}} \cite{ju2024naturalspeech_FACodec} uses FACodec to disentangle speech attributes into subspaces, fused with text tokens from a factorized diffusion model to generate high-quality audio. 

\textbf{USLM} \footnote{\href{https://0nutation.github.io/SpeechTokenizer.github.io/}{https://0nutation.github.io/SpeechTokenizer.github.io/}}  \cite{zhang2023speechtokenizer} builds on SpeechTokenizer to disentangle speech, using autoregressive modeling for content and non-autoregressive modeling for paralinguistic features.

\textbf{SimpleSpeech 1}\footnote{\href{https://simplespeech.github.io/simplespeechDemo/}{https://simplespeech.github.io/simplespeechDemo/}} \cite{yang2024simplespeech_sqcodec} is a non-autoregressive diffusion TTS model using SQ-Codec for compact speech representation. Controls duration via sentence length without alignment. 

\textbf{SimpleSpeech 2} \footnote{\href{https://simplespeech.github.io/simplespeechDemo/}{https://simplespeech.github.io/simplespeechDemo/}} \cite{yang2024simplespeech2} improves SimpleSpeech 1 with SQ-Codec and scalar diffusion, adding multilingual support, better duration modeling, and a time mixture of experts for efficiency. 

\textbf{VALL-E-T}\footnote{\href{https://y-ren16.github.io/TiCodec/}{https://y-ren16.github.io/TiCodec/}} \cite{ren2024fewer_ticodec}, a TiCodec-based VALL-E variant, uses a single quantizer to produce global representations with fewer tokens, reducing word error rates for zero-shot TTS.

\textbf{VALL-E-S}\footnote{\href{https://kkksuper.github.io/Single-Codec/}{https://kkksuper.github.io/Single-Codec/}}  \cite{li2024single}, also a variant of VALL-E, replaces EnCodec with Single-Codec \cite{li2024single}. Single-Codec boosts efficiency and robustness while delivering superior audio quality.

\textbf{MaskGCT}
\footnote{\href{https://huggingface.co/amphion/MaskGCT}{https://huggingface.co/amphion/MaskGCT}}
\cite{wang2024maskgct} is a two-stage TTS with semantic and acoustic codecs. 
It extracts representation from layer 17 of W2v-BERT 2.0 \cite{chung2021w2v}, then uses VQ-GAN \cite{yu2022vectorquantized} and DAC for factorized tokenization. 
The text-to-semantic model employs a masked transformer \cite{touvron2023llama} for in-context learning, while the semantic-to-acoustic model, based on SoundStorm, generates multi-layer acoustic tokens. 
The speech codec, built on DAC, replaces its decoder with Vocos for improved efficiency. 
Audio samples from the repository \cite{amphion} demonstrate high quality. 

\begin{figure}[t]
    \centering
    \includegraphics[width=1.0\columnwidth]{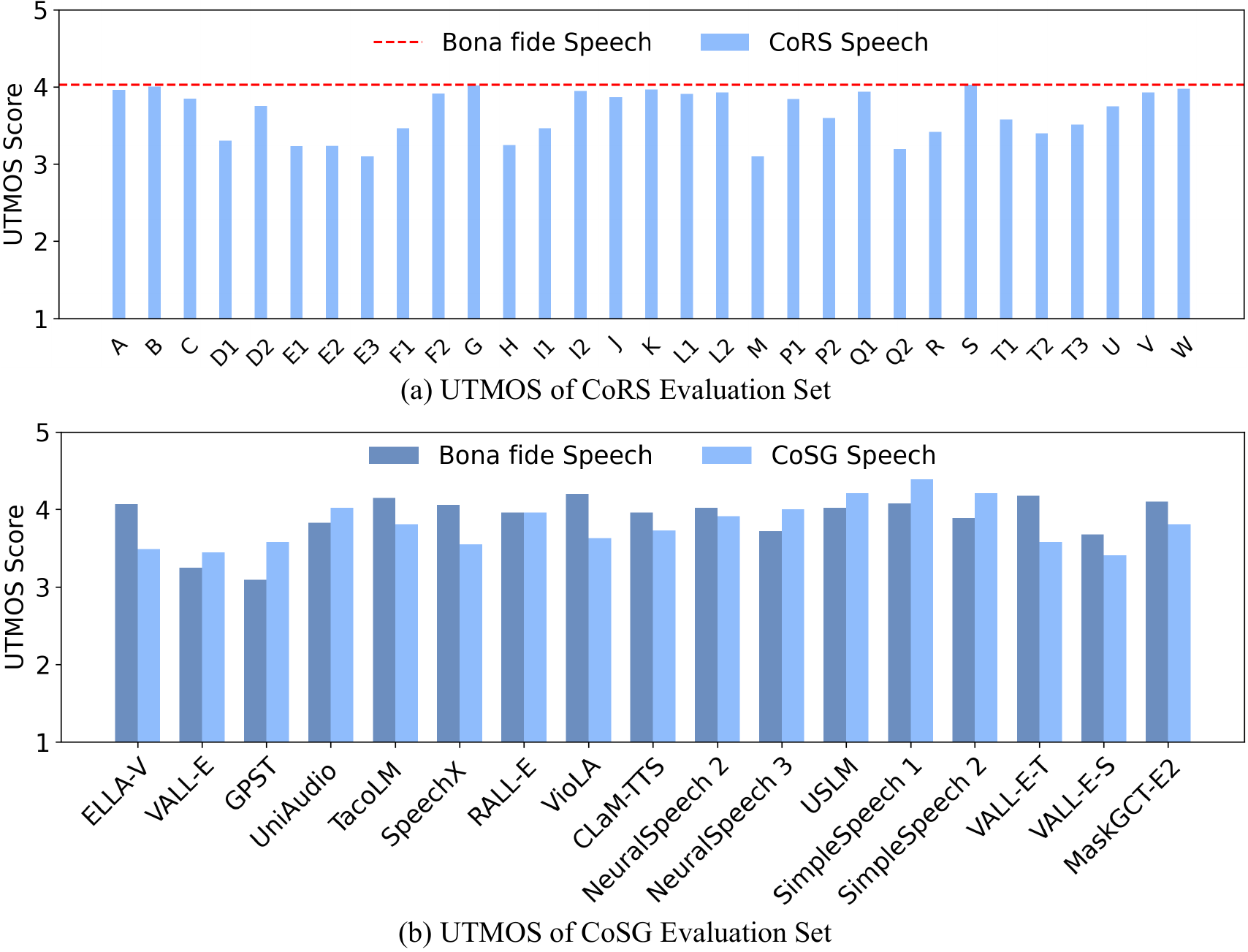}
    \vspace{-17pt}
    \caption{The UTMOS scores of CodecFake+ dataset.}
    \label{fig:codecfake_omni_mos}
\end{figure}

\subsection{Quality Assessment for CodecFake+}\label{lab:quality_assessment}

We use UTMOS \cite{saeki22c_interspeech_utmos}, the top-1 system from VoiceMOS 2022 \cite{huang22f_interspeech_VoiceMOS22}, to predict the mean opinion score (MOS) to approximately assess naturalness. To simplify and emphasize that MOS values are predicted by this pre-trained model, we refer to these scores as UTMOS in this paper. Predicted UTMOS for the CoRS and CoSG sets of CodecFake+ are shown in Figure~\ref{fig:codecfake_omni_mos}.

\subsubsection{CodecFake+ CoRS Evaluation Data}
As shown in Figure~\ref{fig:codecfake_omni_mos}-(a), bona fide speech achieves a UTMOS score of 4.027, which serves as the reference line. Regarding the UTMOS scores of codec re-synthesized speech, DAC, FACodec, and BigCodec have the closest scores to bona fide speech, with values of 4.004, 4.020, and 4.028, respectively.

\subsubsection{CodecFake+ CoSG Evaluation Data} 
Figure~\ref{fig:codecfake_omni_mos}-(b) shows the UTMOS scores of CodecFake+ speech generated by CoSG models. The differences in UTMOS between bona fide and spoofed speech for each CoSG model are less than 0.6, indicating that the CoSG-generated speech (downloaded from their demo page) quality is comparable to bona fide speech. 

\section{Experimental setup}
In this paper, most CMs adopt W2V2-AASIST \cite{tak2022automatic} backbone. 
This choice is driven by its high-capacity SSL representations, which align with Deepfake Arena benchmark~\cite{dowerah2026speech} where leading systems leverage similar SSL foundation model backbones for better generalization. 
Additionally, to showcase the generalization of our datasets, we utilize a lightweight AASIST~\cite{jung2022aasist} model as a baseline to examine whether CoRS data training can also benefit models with low capacity, as detailed in Table~\ref{tab:Cross-Scenario}. 
W2V2-AASIST are trained with RawBoost \cite{tak2022rawboost} as data augmentation, using the development sets to do model selection. 
Models are trained for 4 epochs at the codec level and 20 epochs for all other analyses. 
Other hyperparameters follow the default configurations in \cite{tak2022automatic}: Adam optimizer with a weight decay of $1 \times 10^{-4}$ and an initial learning rate of $1 \times 10^{-6}$, a batch size of 14. 

We employ Equal Error Rate (EER) as the primary metric to assess detection performance, where a lower EER indicates superior accuracy.  
We select the best CMs based on its Pooled EER on the CoRS evaluation set and further assess its generalization on CoSG set.  
To enhance the reliability of cross-scenario and key-factor analyses, we report the pooled EER for the CoSG Eval dataset along with its conditional confidence intervals\footnote{\url{https://github.com/luferrer/ConfidenceIntervals}}, conditioned on the number of CoSG models (Tables~\ref{tab:Cross-Scenario} and~\ref{tab:key_factor}). 
To further analyze taxonomy-level codec artifacts, we report taxonomy-specific average EER (Table~\ref{tab:stratified_analysis}) supplemented by the Standard Error ($SE = SD / \sqrt{n}$), where $n$ denotes the number of subcategory models. This allows us to quantify potential mean variation and facilitate a clearer comparison of model stability. 

We construct our experiments on two databases: CodecFake+ proposed in Section \ref{lab:cf-omni_training_data} and the vocoder-based ASVspoof2019 \cite{wang2020asvspoof}. CodecFake+ CoRS sets are primarily generated based on VCTK. It contains re-synthesized speech (CoRS set) through 31 codec models. It also collects CodecFake speech (CoSG set) generated by 17 CoSG systems. 
As ASVspoof2019 also uses VCTK, we consider it a reference database to explore experiments on the vocoder-based deepfake speech. ASVspoof2019 includes 17 different TTS, VC, and hybrid systems, most of which are vocoder-based. 

\begin{figure*}[ht]
    \centering
    \includegraphics[width=2.0\columnwidth]{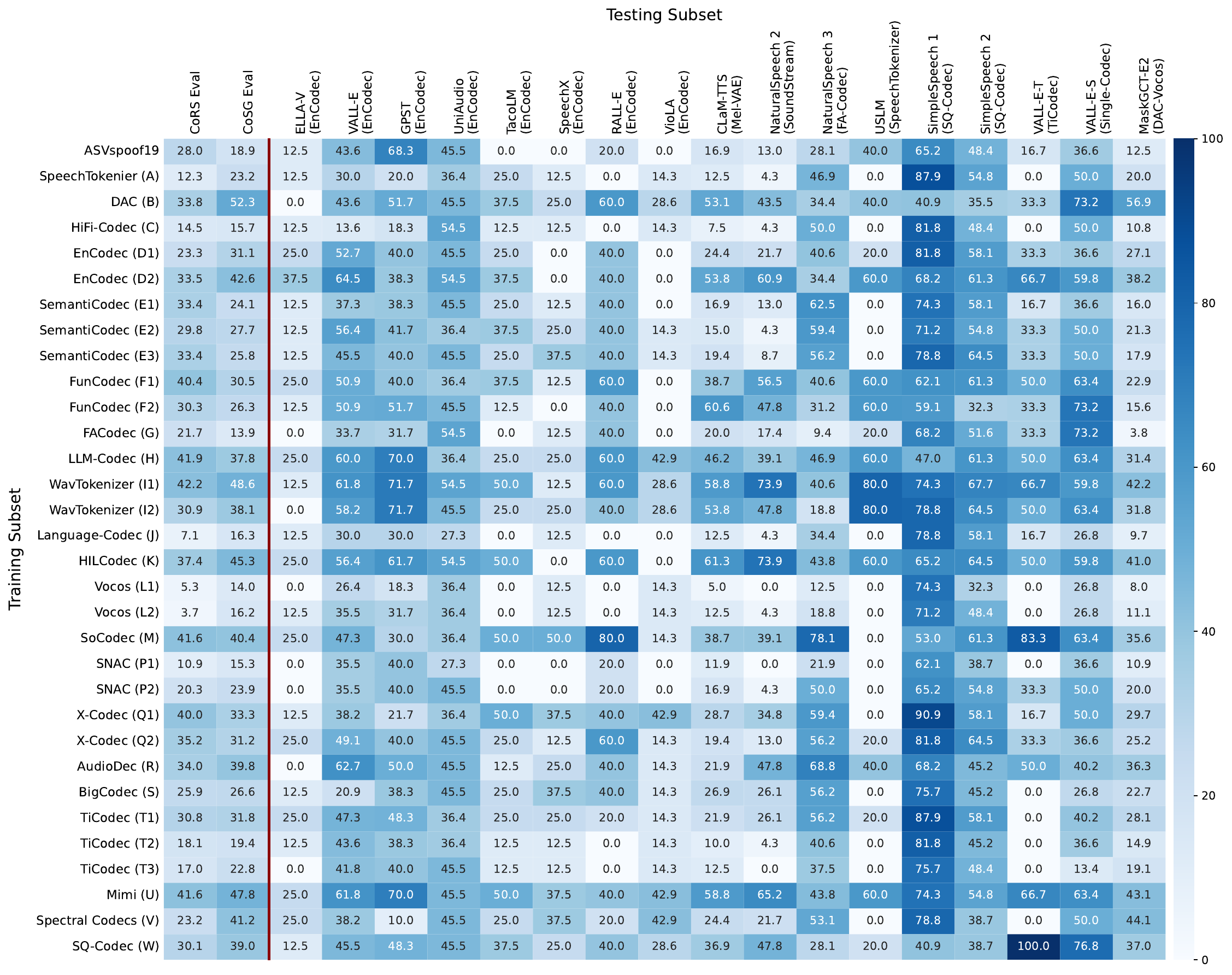}
    \vspace{-7pt}
    \caption{Comparison of CodecFake speech detection performance (EER) for CMs trained on ASVspoof2019 and CoRS dataset (rows A–W). The first column denotes pooled CoRS Eval results, while the second column denotes pooled CoSG Eval results. Other columns are different types of CodecFake Eval subsets.}
    \label{fig:cosg_model_eval}
\end{figure*}

\section{Experimental results}\label{sec:results}
This work aims to comprehensively investigate CodecFake detection through multi-level analysis.  
Codec-level analysis aims to evaluate the effectiveness of CoRS data (of each codec model at the individual codec level) in detecting CodecFake attacks, at the individual codec level.
Taxonomy-level analysis seeks to explore how the intrinsic properties of codecs influence their performance in detecting CodecFake attacks.
Database-level analysis demonstrates the potential of CodecFake+ dataset in defending against CodecFake attacks.

\subsection{Codec-Level Analysis}
\label{subsec:codec-level_analysis}
We first conduct a fine-grained codec-level analysis,
where CMs are trained independently on each codec model (introduced in Table \ref{tab:codec_info}) and evaluated on the CoSG set (introduced in Table \ref{tab:cosg_model}). Results are shown in Fig.~\ref{fig:cosg_model_eval}. The first two columns represent EERs on CoRS and CoSG evaluation sets (overall performance), and the rest of the columns show detailed performance for each CoSG subset.

\subsubsection{How does codec-based re-synthesized data contribute to detecting codec-based deepfake speech?}
To explore this, we first examine the overall performance of CMs trained on different re-synthesized datasets and evaluate them on the CoSG evaluation set of CodecFake+. 
These results are shown in the second column, labeled ``CoSG Eval.,'' in Fig.~\ref{fig:cosg_model_eval}. As a reference, the CM trained on ASVspoof2019 (first row) achieves an EER of 18.9\%.
Comparing CM trained on the ASVspoof2019 with the other rows, six CMs trained on different CoRS training subsets achieve a lower EER, including HiFi-Codec (C), FACodec (G), Language-Codec (J), Vocos (L1-L2), and SNAC (P1).
Notably, models trained on Vocos (L1) and FACodec (G) re-synthesized speech demonstrate superior performance in detecting CoSG speech, achieving a much lower EER below 14\%. 
We have more discussion to determine the key factors for these results in Section \ref{sec:key_factor}.

\subsubsection{What types of CoSG are more difficult to detect?}\label{sec:cosg_difficult_mos}
From the heatmap in Figure~\ref{fig:cosg_model_eval}, we observe that certain CoSG consistently show higher EERs (in dark blue) during detection. 
SimpleSpeech 1 and SimpleSpeech 2 pose significant challenges, as most CMs struggle to detect them. 
This may stem from the high quality of such deepfake speech, which makes detection more challenging. 
This is supported by the high UTMOS scores (indicating potential high naturalness) in Figure~\ref{fig:codecfake_omni_mos}.b, where SimpleSpeech 1 and SimpleSpeech 2 achieve UTMOS scores above 4.15. 
Alternatively, this difficulty may arise from that SimpleSpeech systems use Scq for vector quantization, whereas most CoRS training data are generated using Mvq. 

Lastly, CMs trained on DAC and SQ-Codec perform better on detecting SimpleSpeech 1 and 2. The superior performance of the CM trained on SQ‑Codec likely stems from its matching codec attributes (Scq+None+Time) with those of SimpleSpeech 1 and 2.
This finding further validates the effectiveness of our taxonomy.

\subsection{Taxonomy-Level Analysis}
\label{exp:stratified_analysis}

We observed that good deepfake detection performance is achieved by the similarity in taxonomy between the training (the codecs used in CoRS data) and evaluation (the codecs used in the CoSG systems) data.
We conducted a taxonomy-level analysis based on the taxonomy defeined in Section~\ref{lab:taxonomy}. 
Since deepfake speech with higher UTMOS scores is more challenging to detect as shown in the previous Section~\ref{sec:cosg_difficult_mos}, our analysis focuses on samples with a UTMOS score above 3.5. 
We explore three attributes: vector quantization, auxiliary objectives, and decoder types. 
For evaluating each sub-category (column), we first create an evaluation set containing all data belonging to this sub-category, following Table~\ref{tab:cosg_model}. 
We then organize the CMs (for row) based on the sub-categories used in their respective CoRS training data, following Table~\ref{tab:codec_info}. 
For instance, for the first column of Table~\ref{tab:stratified_analysis}.a, when evaluating CMs on the Svq subset from CoSG Eval., we get an EER for each CM on this Svq subset.
Then, EERs from CMs belonging to the same sub-category (Svq/Mvq/Scq) are averaged separately as the final EER for that sub-category.

\subsubsection{Different Vector Quantizers}\label{sec:res_taxonomy_Quan}
In Table~\ref{tab:stratified_analysis}.a, the EERs along the diagonal are lower than other values in each column. 
This is a different observation compared to Table~\ref{tab:stratified_analysis}.b (auxiliary objectives) and Table~\ref{tab:stratified_analysis}.c (decoder types). 
This indicates that vector quantization has more direct impacts on detection performance than the other two attributes. 
This suggests that using different vector quantization can produce deepfake speech with more distinct distributions. 
Thus, CMs trained on a specific VQ method can effectively detect CoSG data generated using the same VQ method, but may struggle to detect CoSG generated speech with a different VQ.

\begin{table}[t!]
\centering
\caption{Taxonomy-Level Analysis of CodecFake+.}
\label{tab:stratified_analysis}

{ 
\renewcommand{\arraystretch}{1.2} 
\fontsize{7.5}{10}\selectfont
\setlength{\tabcolsep}{4pt}
\newcolumntype{Y}{>{\raggedright\arraybackslash}X}

\newcommand{\conf}[2]{%
    \mbox{#1  {\color{gray}\tiny ($\pm$ #2)}}%
}

\begin{tabularx}{\linewidth}{>{\centering\arraybackslash}m{1.5cm} YYYY}
\toprule

Train & 
\multicolumn{4}{c}{CoSG Eval. Average EER (\%) with Standard Error (SE)}  \\ 
\midrule

\rowcolor{gray!10} \multicolumn{1}{c}{\textit{(a) QUA Schema}} & \textit{Svq} & \textit{Mvq} & \textit{Scq} &  \textit{All} \\
Svq    & \conf{\textbf{16.67}}{16.67$^*$}   & 29.25 &  67.97  & 36.27 \\
Mvq    &  22.62  & \conf{\textbf{24.59}}{4.07}  &  61.39  & \conf{\textbf{28.66}}{3.76} \\
Scq    &  50.00  &  38.77  & \conf{\textbf{48.44}}{8.60}  & 40.06 \\
\midrule

\rowcolor{gray!10} \multicolumn{1}{c}{\textit{(b) AUX Schema}} & \textit{None} & \textit{Sem} & \textit{Disent} &  \textit{All} \\
None   & 28.97 & 20.00 & 35.63 & 31.26 \\
Sem    & 34.04 & 20.00 & 46.49 & 31.36 \\
Disent & \conf{\textbf{21.09}}{5.57} & \conf{\textbf{13.33}}{6.67} & \conf{\textbf{33.33}}{8.37} & \conf{\textbf{25.65}}{5.17} \\
\midrule

\rowcolor{gray!10} \multicolumn{1}{l}{\textit{(c) DEC Schema}} & \textit{Time} & \textit{Freq} & &  \textit{All}\\
Time   &  41.06 &  27.58  & \textcolor{lightgray}{N/A} &  30.48  \\
Freq   & \conf{\textbf{35.45}}{4.42} & \conf{\textbf{21.45}}{6.86} & \textcolor{lightgray}{N/A} & \conf{\textbf{30.22}}{5.94} \\
\bottomrule
\multicolumn{5}{p{0.95\linewidth}}{\scriptsize $^*$ Models I2, S, and T1 were evaluated on a svq subset comprising 6 bona fide and 6 spoofed samples, yielding EERs of 50.00\%, 0.00\%, and 0.00\%, respectively. }\\
\end{tabularx}
} 
\end{table}

\subsubsection{Different Auxiliary Objectives}\label{sec:res_taxonomy_Aux}
In Table~\ref{tab:stratified_analysis}.b, we observe a different pattern compared to Table~\ref{tab:stratified_analysis}.a. 
CMs trained on CoRS data with disentanglement objectives as an auxiliary objective show much lower EERs in the last row, indicating better generalizability in detecting CodecFake speech. 
A possible explanation is that synthetic systems primarily aim to model the speaker's characteristics, content, and prosody. 
Disentanglement codecs re-synthesize speech by disentangling these attributes, ensuring greater control over them. 
Consequently, training with such data enhances the CM model's sensitivity to these attributes, improving CodecFake detection performance. 

\subsubsection{Different Decoders}\label{sec:res_taxonomy_Dec}
In Table~\ref{tab:stratified_analysis}.c, the CM model trained on re-synthesized speech from codecs with frequency-domain decoders (2nd row) achieves lower EERs compared to the CMs trained with data re-synthesized by codec with time-domain decoder (1st row). 
This experiment is just an observation (we don't want to draw a concrete conclusion) due to the imbalance between time- and frequency-domain data.
A balanced training set could lead to more reasonable results and better deepfake detection performance, to be discussed in Table~\ref{tab:Cross-Scenario} (f).

\subsubsection{Take-away}
The reported SE values in Table~\ref{tab:stratified_analysis} confirm that the performance gaps across our taxonomy analysis are statistically meaningful. 
Our taxonomy-level analysis highlights: 1) Using an aligned vector quantization is crucial for improving performance on the evaluation set generated by CoSG with the same vector quantization. 2) Disentanglement auxiliary objectives can enhance generalizability in detecting deepfake speech generated by other methods. 

\subsection{Database-Level Analysis}
\subsubsection{Training Data Selection Strategies}

With diverse codecs, training a CM on a pooled dataset intuitively enhances generalization. 
However, when we trained a CM on the entire pooled CoRS Train set using random sampling for balanced CoRS Train sets, EER on CoSG Eval. reached 50\%, indicating overfitting. 
A similar issue was reported in the copy-synthesis method (Appendix A.2.2 of \cite{wang2023spoofed}) designed for vocoder-based databases, even when using advanced techniques. 
This suggests that simply aggregating diverse CoRS data makes it difficult for the CM to generalize effectively. 
Addressing this challenge requires a more strategic approach to data selection.

\noindent \textit{\textbf{Performance-Driven Selection}}.
A straightforward way is to select data used by the top-performance CMs. 
Thus, we selected CoRS subsets (J, L1, and L2) corresponding to the top three performing CM models (Figure~\ref{fig:cosg_model_eval}) to train a new CM under different backbones. 
The results are in rows (c-e) and (h-j) of Table~\ref{tab:Cross-Scenario}. 
We can see that this new CM, trained on selected subsets (J, L1, and L2 subsets), mitigates the overfitting observed when using the entire CoRS dataset, highlighting the importance of strategic data selection. 
Furthermore, combining the selected CoRS subsets with existing datasets (ASVspoof2019 or ASVspoof5) could significantly improve CoSG performance while maintaining robustness on the ASVspoof2019 evaluation set. 
Notably, when utilizing the W2V2-AASIST backbone, the combination of CoRS and ASVspoof5 (row j) achieves the best performance on the CoSG evaluation set with a Pooled EER of 8.40\%. 
This breakthrough indicates a strong complementarity between CoRS and various vocoder-based data for detecting CodecFake speech, and demonstrates that pre-trained features (W2V2) further enhance the model's ability to generalize across unseen codecs.

\noindent\textit{\textbf{Taxonomy-Guided Balancing}}.
To achieve a more effective balance in training data, we propose a taxonomy-guided selection approach based on the taxonomy defined in Section~\ref{lab:taxonomy}. 
This includes three balanced sampling strategies: QUA balance, AUX balance, and DEC balance. Each ensures equal distribution across the corresponding sub-categories. 
For example, in DEC balance, the number of training CoRS data utterances derived from time-domain (Time) and frequency-domain (Freq) decoders is similar.
The results are shown in the last three rows of Table \ref{tab:Cross-Scenario} (k)-(m). 
Interestingly, training on the full CoRS data causes overfitting, while taxonomy-guided balanced selection improves performance. 
This further highlights the value of our taxonomy, providing guidance on leveraging the large CodecFake+ dataset to get better performance. 
In addition, the DEC balance even achieves the lowest EER (11.91\%) on the CoSG Eval. 
The DEC balance strategy excels by forcing the model to learn discriminative decoder-domain features. As detailed in Section~\ref{sec:key_artifact_factor}, decoder types define the structural identity of artifacts with clearer morphological distinctions than auxiliary objectives. 

\noindent \textit{\textbf{Training Data Scaling.}}
To evaluate the impact of data volume, we increased the DEC-balanced proxy data to four times ($\times 4$) its original size. 
As shown in Table \ref{tab:Cross-Scenario} (n), this expansion further lowers the pooled EER on CoSG Eval to 10.85\%. 
However, this performance gain does not inherently resolve the instability issue, as evidenced by the persistently wide confidence interval. This suggests that while scaling training volume improves average accuracy, it is insufficient on its own to ensure robust generalization across disparate scenarios. 

\begin{table}[t]
\centering
\caption{Database-Level Cross-Scenario Analysis.}
\label{tab:Cross-Scenario}
\fontsize{7.5}{10}\selectfont 
\renewcommand{\arraystretch}{1.2} 
\setlength\tabcolsep{4.5pt} 

\begin{tabular}{ll c l}
\toprule
\multirow{2}{*}{\textbf{ID}} & \multirow{2}{*}{\textbf{Training Set}} & \multicolumn{2}{c}{\textbf{Pooled EER (\%)} $\downarrow$} \\
\cmidrule(lr){3-4}
& & {\textbf{19LA Eval.}} & {\textbf{CoSG Eval.}} \\
\midrule

\rowcolor{gray!10} \multicolumn{4}{l}{\textit{Traditional vs. Proxy-Augmented Models (Backbone: AASIST)}} \\
(a) & ASVspoof19  & 15.90 & 36.95 {\tiny (33.46--46.62)} \\
(b) & ASVspoof5    & 26.57 & 29.43 {\tiny (23.46--41.60)} \\
\cmidrule(r){2-4}
(c) & CoRS (J, L1, L2) & 13.89 & 24.60 {\tiny (20.92--36.10)} \\
(d) & CoRS (J, L1, L2) + ASVspoof19 & \phantom{0}\textbf{2.73} & 29.77 {\tiny (26.29--36.62)} \\
(e) & CoRS (J, L1, L2) + ASVspoof5  & 26.57 & \textbf{24.59} {\tiny (17.91--27.80)} \\

\midrule

\rowcolor{gray!10} \multicolumn{4}{l}{\textit{Traditional vs. Proxy-Augmented Models (Backbone: W2V2-AASIST)}} \\
(f) & ASVspoof19 & {\phantom{0}\textbf{0.12}} & 18.92 {\tiny (13.89--38.64)} \\
(g) & ASVspoof5  & 18.03 & 21.81 {\tiny (17.16--37.14)} \\
\cmidrule(r){2-4}
(h) & CoRS (J, L1, L2) & \phantom{0}1.10    & 14.09 {\tiny \phantom{0}(8.65-41.41)} \\
(i) & CoRS (J, L1, L2) + ASVspoof19 & \phantom{0}0.53 & {12.97} {\tiny \phantom{0}(7.82-39.11)} \\
(j) & CoRS (J, L1, L2) + ASVspoof5 & \phantom{0}8.53 & \phantom{0}\textbf{8.40} {\tiny \phantom{0}(2.71--29.43)} \\

\midrule

\rowcolor{gray!5} \multicolumn{4}{l}{\textit{Data Balance and Scaling (Backbone: W2V2-AASIST)}} \\
(k) & CoRS (QUA Balance) & \phantom{0}1.93 & 21.93 {\tiny (16.62--28.01)} \\
(l) & CoRS (AUX Balance) & \phantom{0}2.18 & 15.02 {\tiny (11.07--28.35)} \\
(m) & CoRS (DEC Balance) & \phantom{0}\textbf{1.51} & {11.91} {\tiny \phantom{0}(6.93--34.01)} \\
\cmidrule(r){2-4}
(n) & CoRS (DEC Balance $\times$4) & 2.27 & 10.85 {\tiny \phantom{0}(4.91--35.63)} \\
(o) & CoRS (DEC Balance $\times$4) + ASVspoof5 & 2.72 & \phantom{0}\textbf{8.07} {\tiny \phantom{0}(3.32--23.18)} \\
\bottomrule
\end{tabular}
\end{table}

\subsubsection{Cross-Scenario Analysis}\label{sec:database_level_cross_scenario}
Based on the two proposed data selection strategies, we conduct a cross-scenario analysis to investigate \textit{whether a model trained on codec-based data can generalize to vocoder-based data and vice versa}.  
Table \ref{tab:Cross-Scenario} lists the results, with conditional confidence intervals added to the CoSG Eval scores to measure performance stability. 

First, we evaluate the limitations of existing datasets. Note that the ASVspoof 5 evaluation set is excluded because we use different labeling under a different motivation, as explained in Section~\ref{sec:codecfake_definition}.
Consistent with prior findings \cite{wu24p_interspeech}, the ASVspoof2019-trained baseline fails to generalize to CodecFake detection. Even a model trained on the more diverse ASVspoof5 dataset exhibits poor generalization across both 19LA and CoSG evaluation sets compared with the model in row (c) trained on the CoRS subset. This underscores that simply increasing data volume or diversity is insufficient to bridge the gap between traditional vocoder-based artifacts and emerging codec-based threats. 

Second, regarding the proposed strategies, QUA balance (k) improves CoSG detection and provides a narrower confidence interval, though it leads to a slight EER increase on 19LA. DEC balance (m) achieves a lower pooled EER of 11.91\% on CoSG, but exhibits higher variance in its performance.

Finally, The limitations of scaling lead us to explore joint training with the ASVspoof5 dataset to optimize cross-scenario reliability. This combination (Row o) achieves the optimal EER of 8.07\% and, more importantly, yields a significantly tighter confidence interval (3.32\%--23.18\%). These results demonstrate that the taxonomy-guided codec data and traditional vocoder-based data complement each other.

In summary, our results suggest that solving the generalization puzzle requires a shift from ``quantity-driven" to ``artifact-aware" data selection. Although cross-scenario detection persists, the synergy observed between diverse datasets \cite{zhu24_asvspoof, marek2024audio, li2025we} offers a promising path forward. We believe these insights provide a crucial roadmap for building the next generation of robust, generalized deepfake detection systems.

\begin{figure*}[t]
    \centering
    \includegraphics[width=2\columnwidth]{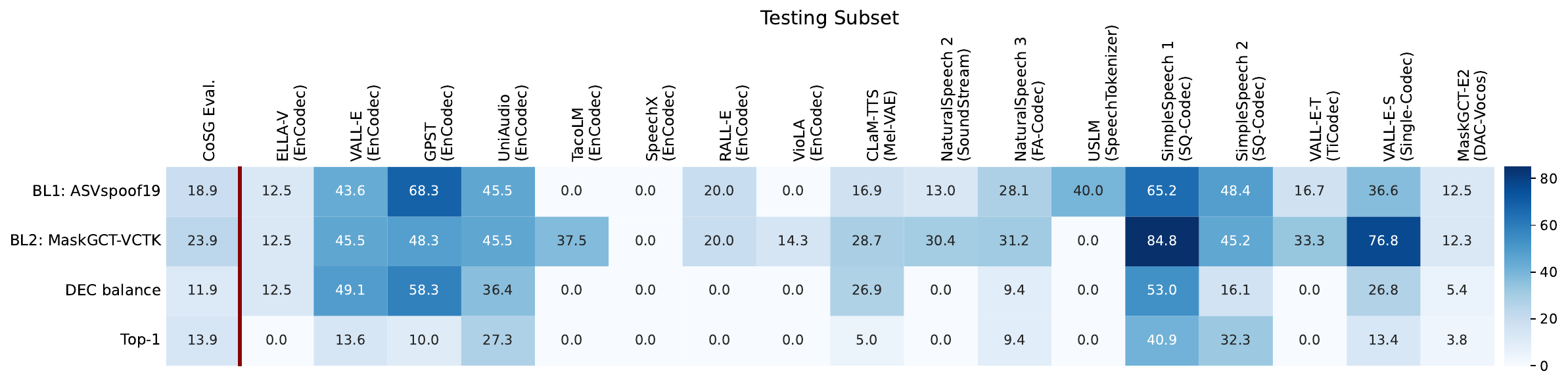}
    \vspace{-7pt}
    \caption{\mrev{}{Pooled EERs (\%)} of CMs trained on ASVspoof2019, MaskGT-VCTK, and DEC balance, the selected CoRS subsets, evaluated on CodecFake CoSG Eval. The y-axis, `Eval. Subset (Codec)' presents the CoSG Eval subset and the codec. 
    \textbf{Top-1} denotes the best CM using CoRS for training (based on Fig. \ref{fig:cosg_model_eval}).}
    \label{fig:top1_eval}
\end{figure*}

\section{Discussion}
Section \ref{sec:results} focuses on training CMs with re-synthesized CoRS speech, since most CoSG models are unavailable due to license concerns. 
Here, we consider training CMs with speech from a specific CoSG model instead of CoRS to detect fake speech from other unseen CoSG models. 
We use MaskGCT-VCTK to generate training CoSG data as a controlled testbed. Its hybrid architecture (DAC encoder with Vocos decoder) allows us to isolate codec-specific factors from content variations when using identical source audio (VCTK). 
Then, we use our taxonomy to pinpoint which codec factors produce better CoRS training data.

\subsection{Is CoSG Data Better than CoRS Data for Training?}
To evaluate the effectiveness of CoSG versus CoRS for training, we compare CMs trained on different datasets in Figure~\ref{fig:top1_eval}. 
The first three rows are distinguished by their training data: ASVspoof 2019, MaskGCT-VCTK, and DEC Balance. 
The last row shows the CM selected based on the top-1 performance (Figure \ref{fig:cosg_model_eval}) for each CoSG dataset (column).
The first column of Figure~\ref{fig:top1_eval} shows the pooled CoSG evaluation set, while the subsequent columns present the breakdown of EERs for each CoSG method. 

In the ``CoSG Eval.'' column, CM trained on MaskGCT-VCTK (BL2) achieves a relatively higher EER than others. 
In the last column, MaskGCT-E2, it still performs much worse than Top-1 with EER=12.3\%, even though both training and testing data were generated by the MaskGCT. 
The main difference is the source data used by MaskGCT for data generation: VCTK is used for the training set generation, and the E2 TTS test set is used for generating the testing set. 
BL2 might be overfitted to the training data sourced from the VCTK dataset, limiting its ability to generalize effectively to other types of CoSG models and other kinds of source datasets. 

In addition, when examining the EER breakdown across different CoSG subsets (2nd column to the end), we find that 
the Top-1 CM, trained on a subset of CoRS, consistently outperforms both BL1 and BL2. 
DEC Balance outperforms the CM trained on MaskGCT-VCTK when evaluated on most CoSG subsets.
These findings demonstrate the effectiveness of using CoRS data to simulate CoSG data when they are not readily available 
and emphasize the potential of the CodecFake+ dataset in improving CodecFake detection performance. 
Despite its architectural uniqueness, limited CoSG for training may introduce overfitting due to limited diversity.
Our results with MaskGCT-VCTK further confirm that relying solely on non-diverse CoSG systems to generate training data may result in models with poor generalization.

\begin{table}[t]
\centering
\fontsize{7.5}{10}\selectfont 
\renewcommand{\arraystretch}{1.2} 
\setlength\tabcolsep{5pt}
\caption{Key Factor Analysis of CoRS Speech (EER \%).}
\label{tab:key_factor}

\begin{tabularx}{\linewidth}{l ll cc}
\toprule
& \multicolumn{2}{c}{Training Configuration} & \multicolumn{2}{c}{Pooled EER (\%) $\downarrow$} \\
\cmidrule(lr){2-3} \cmidrule(lr){4-5}
ID & Name & Data Type & MaskGCT-VCTK & MaskGCT-E2 \\
\midrule
\rowcolor{gray!10} \multicolumn{5}{l}{\textit{Traditional Baselines}} \\
BL1 & ASVspoof 2019 & Traditional & 27.95 {\tiny (21.27--34.04)} & 12.50 {\tiny (10.55--14.60)} \\
BL2 & MaskGCT-VCTK & CoSG & 0.00 {\tiny (0.00--0.00)} & 12.33 {\tiny (10.26--14.60)} \\
\midrule
\rowcolor{gray!10} \multicolumn{5}{l}{\textit{CoRS Proxies with Shared Codec Components}} \\
D1 & EnCodec & CoRS & 12.85 {\tiny \phantom{0}(5.90--20.21)} & 27.08 {\tiny (24.71--29.97)} \\
D2 & EnCodec & CoRS & 37.22 {\tiny (32.41--41.09)} & 38.19 {\tiny (35.63--41.40)} \\
B  & DAC     & CoRS & 22.52 {\tiny (10.70--34.80)} & 56.94 {\tiny (54.19--59.75)} \\
L1 & Vocos   & CoRS & 0.40 {\tiny (0.00--1.03)}    &  7.99 {\tiny (6.41--9.58)}  \\
L2 & Vocos   & CoRS & 0.26 {\tiny (0.00--0.81)}    & 11.11 {\tiny (8.73--13.14)} \\
\midrule
\rowcolor{gray!10} \multicolumn{5}{l}{\textit{Best-Performing CoRS Proxy (Top-1 in Fig.~\ref{fig:top1_eval} )}} \\
G & FACodec & CoRS &  22.78 {\tiny (15.00--30.71)}  & \phantom{0}\textbf{3.82} {\tiny (2.68--5.02)} \\
\bottomrule
\end{tabularx}
\end{table}

Although CoRS doesn't cover all codecs used in the testing set, it shows generalizability when evaluating the unseen CoSG speech with an unseen codec. 
For example, the top-1 CMs can detect CoSG speech generated by CLaM-TTS (Mel-VAE), NaturalSpeech 2 (SoundStream), VALL-E-S (Single-Codec), and MaskGCT-E2 (DAC-Vocos), even though their codec are not shown in the CoRS training set. 
The best CM trained on DEC Balance dataset exhibits similar trends. For example, the DEC Balance-trained CM can effectively detect CoSG speech generated by NaturalSpeech 2 (SoundStream), VALL-E-S (Single-Codec), and MaskGCT-E2 (DAC-Vocos), even though their codecs are not present in the CoRS training set. 

These results in Fig.~\ref{fig:top1_eval} show that training with diverse CoRS simulation is significantly more effective than relying on target-specific CoSG data. Even when trained on data generated by the exact target model (MaskGCT), the CM suffers from severe overfitting (23.9\% EER). In contrast, training on data from a single codec proxy (FACodec) reduces the EER to 13.9\%, while the DEC-Balance strategy further optimizes it to 11.9\%. This confirms that high-diversity CoRS is a more reliable training source than model-specific CoSG, providing superior generalization to unseen codecs and architectures. 

\subsection{Which Factors of CoRS Contribute to Better Performance?}\label{sec:key_factor}
Table~\ref{tab:key_factor} presents key CoRS factors contributing to CM performance. 
In Table~\ref{tab:key_factor}, we first investigate how CoRS training data with similar components (with codecs used in CoSG systems) can improve final performance and further explore the key influencing factors. 
We evaluate CMs trained on CoRS subsets from EnCodec, DAC, and Vocos.
These codecs share components with MaskGCT. 
MaskGCT builds on DAC, modifying discriminators and losses, and replacing the decoder with one aligned to Vocos.  
As a result, both DAC and Vocos are included. 
As DAC extends EnCodec with advanced training, EnCodec is also included.
Furthermore, as the CM trained on the CoRS subset generated by FACodec achieves the top-1 performance, FACodec is also included. 

\begin{figure*}[t]
\centering
\includegraphics[width=2\columnwidth]{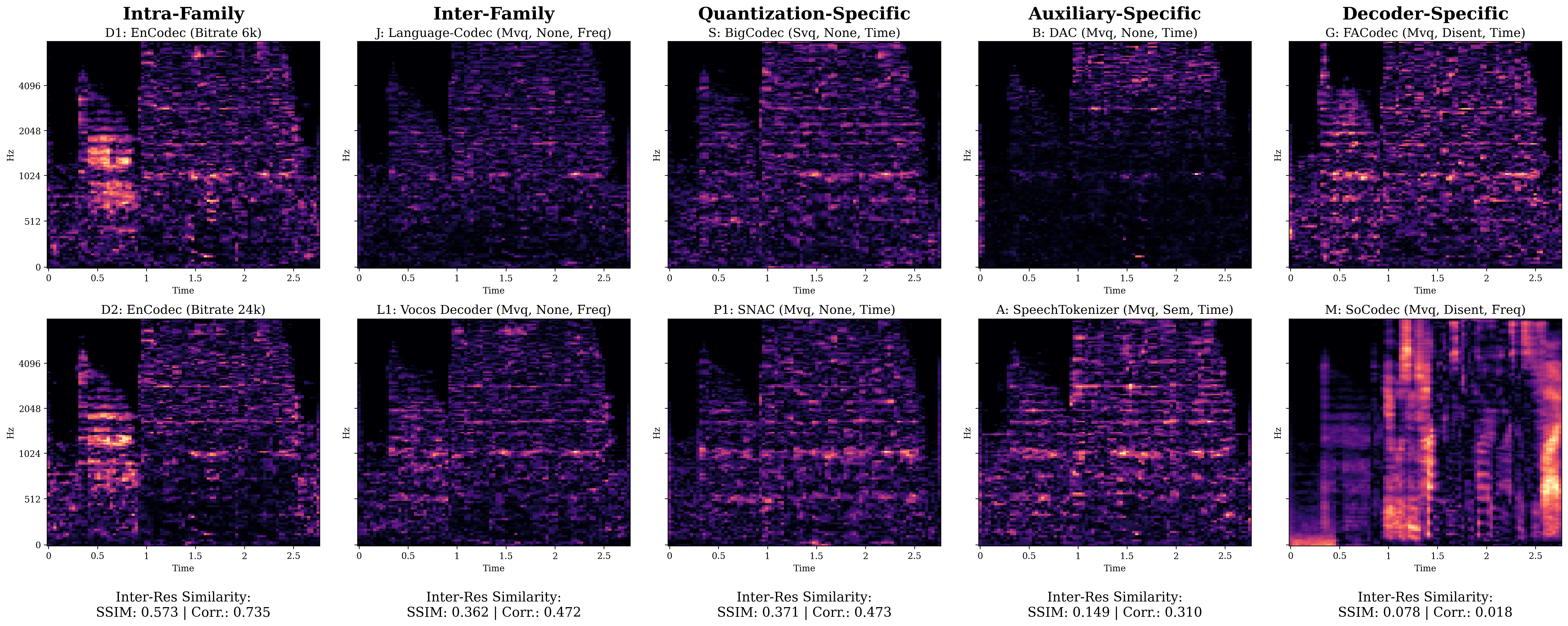}
\vspace{-7pt}
\caption{Impact of various factors on neural audio codec artifacts. We analyze these artifacts using residual spectrograms, which capture the difference between original and CoRS audio. Across the five groups, we observe a progressive shift: intra-family, inter-family, and quantization-specific codecs exhibit high structural similarity, whereas auxiliary- and decoder-specific variations show more codec structural artifacts distinction.}
\label{fig:spec_visual}
\end{figure*}

\begin{enumerate}[itemsep=0mm, leftmargin=4mm, nosep]
\item Compared to the CM trained on MaskGCT-VCTK (2nd row), CMs trained on EnCodec (D1-D2) and DAC (B) achieve worse EERs despite sharing similar codec encoders. 
This hints that training and testing data with similar codec encoder types do not contribute much to the performance. 
On the other hand, CMs trained on Vocos (L1-L2) outperform the CM trained on MaskGCT-VCTK. 
Both the codec used in MaskGCT and Vocos codec employ a frequency-domain decoder, and this similarity in decoder type plays a crucial role in enhancing detection performance. 
This observation aligns with our analysis in Section~\ref{sec:res_taxonomy_Dec}.

\item CM trained on CoRS training data re-synthesized by FACodec achieves the best performance in this comparison. 
FACodec utilizes a disentanglement auxiliary objective to separate content, speaker, and prosody information. 
The superior detection performance of the FACodec subset suggests that explicitly disentangling these speech attributes may improve CodecFake detection. 
Even though MaskGCT does not incorporate disentanglement components, the CM trained on FACodec performs pretty good. 
This indicates that CoRS data using a disentanglement-based codec can enhance the generalizability of the CM model.
This conclusion also aligns with our findings in Section~\ref{sec:res_taxonomy_Aux}.
\end{enumerate}
 
In summary, the frequency decoder and disentanglement are factors that may contribute to generalization for CodecFake detection, enabling CM to perform well on unseen CoSG data. 

We compared CMs trained on CoRS versus CoSG data and asked: 
(1) Does training on CoSG data yield better results than using CoRS as a proxy? Our findings suggest that while CoSG data is the direct target, relying on a single or limited set of CoSG systems for training leads to poor generalization. Specifically, the CM trained on MaskGCT-VCTK fails to generalize even to the same system on a different source dataset (MaskGCT-E2). In contrast, CoRS provides a more \textbf{scalable and diverse} training source that captures generalized codec-induced artifacts, making it a superior alternative when diverse CoSG models are inaccessible. 
When the CoSG data is limited, it tends to overfit specific data types (source data and CoSG models). Whereas, CoRS data can provide broader generalization across diverse CoSG systems.
(2) Which CoRS factors drive performance? We evaluated CMs on CoRS subsets sharing components with the CoSG test set.  
Results show performance hinges on shared codec components. In particular, frequency-domain decoders and disentanglement mechanisms are key factors that contribute to better generalizability. 

\subsection{What Governs Structural Differences in Codec Artifacts?} \label{sec:key_artifact_factor}
We analyze the structural differences in codec artifacts through residual spectrograms, as illustrated in Fig.~\ref{fig:spec_visual}.
Following the taxonomy in Section~\ref{lab:taxonomy}, we investigate the inductive biases behind forensic artifact formation. We use residual spectrograms (the residual between original spectrograms and reconstructed spectrograms) to quantify how various codec configurations reshape forensic textures.
To systematically evaluate these effects, we analyze five perspectives:  
\begin{itemize} 
\item \textbf{Intra-Family Consistency:} Assesses sensitivity to configuration variations (e.g., bitrates, training data) within the same codec taxonomy. 
\item \textbf{Inter-Family Consistency:} Evaluates structural convergence across different codec families within the same taxonomic subcategory. 
\item \textbf{Quantization-Specific Discrepancy:} Isolates the impact of vector quantization while maintaining consistent taxonomic attributes elsewhere.
\item \textbf{Auxiliary-Specific Discrepancy:} Investigates the impact of auxiliary objectives by keeping other neural codec taxonomic attributes consistent.
\item \textbf{Decoder-Specific Discrepancy:} Measures structural divergence caused by varying the reconstruction domain (e.g., time vs. frequency) while maintaining a consistent taxonomy for all other components. 
\end{itemize} 
We further utilize SSIM and Pearson Correlation (Corr) for pairwise comparisons across 100 CoRS samples. The results are shown in Table~\ref{tab:structural_stat}. Higher scores denote structural invariance, while lower scores signify artifact dissociation.

As illustrated by the representative samples in Fig.~\ref{fig:spec_visual}, residual spectrograms reveal a progressive shift from structural cohesion to divergence. 
While the initial groups with taxonomy consistency display consistent spectral ``fingerprints'' with overlapping artifact patterns, the latter groups with taxonomy discrepancy variations undergo a morphological difference. 
Quantitative comparisons in Table \ref{tab:structural_stat} confirm these observations. 
Taking SSIM as a primary indicator, the \texttt{Intra-Family} and \texttt{Inter-Family} groups maintain high similarity with SSIM values between 0.4483 and 0.3122, indicating that the core forensic signature remains similar. 
Surprisingly, the \texttt{Quantization-Specific} group yields an SSIM of 0.3225, suggesting that quantization discrepancies do not fundamentally alter the artifact structure visible in the spectrograms. 
In contrast, the \texttt{Auxiliary-} and \texttt{Decoder-Specific} variations exhibit a structural distinction, where SSIM metrics reach a nadir of 0.2215 and 0.2520. Another metric Pearson Correlation, follow a consistently decline. 
These results demonstrate that while factors such as bitrates primarily affect artifact intensity, the decoder types and auxiliary objectives are the primary determinants of the structural identity of neural audio codec artifacts. 

\section{Conclusion}

To counter and better understand the emerging CodecFake, we present CodecFake+, a large-scale dataset reflecting its rapid progress.
As of February 2025, CodecFake+ is, to our knowledge, the largest publicly available dataset in this domain, covering 31 open-source neural codecs and CodecFake speech from 17 advanced CoSG systems.
Our primary contribution is a unified neural audio codec taxonomy that categorizes systems by vector quantizers, auxiliary objectives, and decoder types, enabling a structural understanding of how architectural choices govern forensic artifacts.
We then conduct multi-level experiments spanning codec, taxonomy, and database perspectives, showing that CoRS data effectively improves detection of CoSG speech by providing a more scalable and diverse training source that captures generalized codec-induced artifacts.
Notably, our taxonomy reveals that shared codec components, particularly frequency-domain decoders and disentanglement mechanisms, provides a useful basis for constructing efficient training datasets. Our structural artifact analysis further demonstrates that the decoder type and auxiliary objectives are the primary determinants of the forensic identity in neural audio codecs.

\begin{table}[t]
\centering
\fontsize{8.5}{11}\selectfont 
\renewcommand{\arraystretch}{1.2} 
\setlength\tabcolsep{8.5pt} 
\caption{Structural Artifact Analysis across Different Groups.}
\label{tab:structural_stat}

\begin{tabularx}{\linewidth}{l cc}
\toprule
Group Name & Avg. SSIM & Avg. Corr \\
\midrule
\rowcolor{gray!10} \multicolumn{3}{l}{\textit{Neural Audio Codec Taxonomy Consistency}} \\
Intra-Family Taxonomy Consistency & 0.4483 & 0.5245 \\
Inter-Family Taxonomy Consistency & 0.3122 & 0.3646 \\
\midrule
\rowcolor{gray!10} \multicolumn{3}{l}{\textit{Neural Audio Codec Taxonomy Discrepancy}} \\
Quantization-Specific Discrepancy & 0.3225 & 0.3868 \\
Auxiliary-Specific Discrepancy & 0.2520 & 0.2963 \\
Decoder-Specific Discrepancy   & 0.2215 & 0.2406 \\
\bottomrule
\end{tabularx}
\end{table}

\section{Limitations and Future Work}
This paper establishes a taxonomy for codec-based deepfakes, though limitations regarding evaluation metrics, methodological improvements  and data diversity persist.

\textbf{Evaluation Metric.} While UTMOS is a widely used automatic assessment metric for estimate quality, its sensitivity to unique neural codec artifacts remains an open question compared to traditional acoustic errors. Furthermore, full-reference metrics such as PESQ are inapplicable because CoSG models lack time-aligned ground truth. Following recent neural audio codec survey and benchmarks \cite{mousavi2025discrete}, we use UTMOS metric to analyze relative score gaps rather than absolute values to mitigate evaluation metric bias. 

\textbf{Methodological Improvements.}
Although taxonomy-guided subset selection for CoRS yielded strong results, further refinement in strategic training may further enhance performance. Additionally, while frequency-domain decoders and disentanglement mechanisms appear to boost generalization, deeper investigation is needed to isolate their specific mechanisms. As the field evolves, we hope CodecFake+ provides the foundation for solving these remaining challenges.

\textbf{Data Diversity and Robustness.} Currently, CodecFake+ dataset focuses on studio-quality English speech from the VCTK dataset. We plan to leverage our codec taxonomy to expand the benchmark into three key areas: (1) multilingual datasets to verify language independence; (2) noisy, in-the-wild recordings to evaluate robustness; and (3) spontaneous conversational speech. This research roadmap ensures that codec-based countermeasures generalize effectively across diverse real-world acoustic environments.

\bibliographystyle{IEEEtran}
\bibliography{refs}

\end{document}